\documentclass{jfm}
\usepackage{graphicx}
\usepackage{xcolor}
\usepackage{booktabs,longtable,array,pdflscape}
\usepackage{capt-of}
\usepackage{amsmath}
\newcommand{\redtext}[1]{{\color{black}#1}}
\newcommand{\bluetext}[1]{{\color{black}#1}}

\renewcommand{\eqref}[1]{(\ref{#1})}
\shorttitle{Bridging the local and the global}
\shortauthor{Y.-S. Zhang, Y.-F. Li, M.-J. Xiao and Y.-H. Wang}
\title{Bridging the local and the global: a physically constrained buoyancy--drag model for unified prediction of Rayleigh--Taylor and Richtmyer--Meshkov mixing widths across density ratios}
\author{You-Sheng Zhang\aff{2,3,4},
Ya-Feng Li\aff{1}\corresp{Email address for correspondence: 1192373799@qq.com},
Meng-Juan Xiao\aff{2,3}
\and
Yu-Hui Wang\aff{1}\corresp{Email address for correspondence: aowuki@163.com}}
\affiliation{\aff{1}College of Mechanical and Electrical Engineering, Beijing University of Chemical Technology, Beijing 100029, China
\aff{2}Institute of Applied Physics and Computational Mathematics, Beijing 100094, China
\aff{3}National Key Laboratory of Computational Physics, Beijing 100088, China
\aff{4}Center for Applied Physics and Technology, HEDPS, College of Engineering, Peking University, Beijing 100871, China}
\begin{document}
\maketitle
\begin{abstract}
Accurate prediction of the macroscopic width of Rayleigh--Taylor (RT) and Richtmyer--Meshkov (RM) turbulent mixing layers is central to inertial confinement fusion and supernova dynamics. However, bubble--spike asymmetry, density-ratio dependence and unsteady forcing pose a persistent closure challenge: existing low-order buoyancy--drag models struggle to describe different mixing problems accurately with one model and coefficient set. We combine local front dynamics with global mass conservation in separate buoyancy--drag equations for the bubble and spike fronts. Rather than imposing shared or fixed empirical coefficients, the model retains separate inertia, buoyancy and drag coefficients on the two sides and allows them to vary independently with density ratio. Given the bubble-side state scalings, RT/RM similarity relations, a mean-composition profile and endpoint asymptotics jointly constrain all six effective coefficients without case-by-case fitting. A profile-shape parameter $c$ labels distinct internal composition states and is selected a priori from RT spike scaling measured in linear-electric-motor experiments. The model then cross-predicts the RM spike exponent without recalibration to RM spike data and, by construction, recovers low-Atwood-number bubble--spike symmetry and the high-density-ratio free-fall RT-spike and ballistic RM-spike limits. Tests against constant- and variable-acceleration RT mixing, post-impulse RM evolution and Nova laser deceleration show that one closure describes mixing-width evolution across density ratios and acceleration histories without case-specific retuning, while reducing excessive spike growth at high density ratio. This physically interpretable, asymptotically consistent framework enables cross-problem prediction of wide-density-ratio RT and post-impulse RM mixing.
\end{abstract}

\section{Introduction}\label{sec:1}

When a light fluid supports or accelerates a heavy fluid, perturbations of their interface become unstable and develop into Rayleigh--Taylor (RT) turbulent mixing \citep{rayleigh1883,taylor1950}. A shock crossing a perturbed material interface likewise amplifies the disturbance and can generate Richtmyer--Meshkov (RM) turbulent mixing \citep{richtmyer1960,meshkov1969}. RT and RM mixing govern material transport in systems ranging from supernova explosions and inertial-confinement fusion to shock-driven reactive fuel--air interfaces, where interfacial instability interacts with ignition dynamics \bluetext{\citep{zhou2017,hillebrandt2000,remington2006,lindl1995,liu2026shock}}. In fusion capsules, unstable interfacial mixing carries cold dense material into the hot spot, degrades its energy balance and can determine whether ignition is achieved. In supernovae, it redistributes chemical elements and shapes the observable ejecta. Predicting the evolution of the mixing region is therefore important both fundamentally and technologically \citep{hillebrandt2000,remington2006,lindl1995}.

The mixing width is the most widely used integral measure of this evolution \bluetext{\citep{liu2025bubble,zhangni2023bubblemerge,read1984,dimonteschneider2000,youngs2013,zhou2017}}. We denote by $h_s$ the spike width, over which heavy fluid penetrates the light fluid, and by $h_b$ the bubble width, over which light fluid penetrates the heavy fluid; the total width is $h=h_b+h_s$. The principal control parameters are the acceleration history $g(t)$ and the density ratio $R\equiv\rho_2/\rho_1$, or equivalently the Atwood number $A\equiv(R-1)/(R+1)\in[0,1]$, where $\rho_1$ and $\rho_2$ are the light- and heavy-fluid densities. In the self-similar regime, classical RT mixing follows $h_{b,s}=\alpha_{b,s}Agt^2$, whereas post-impulse RM mixing follows $h_{b,s}\propto t^{\theta_{b,s}}$ \citep{dimonteschneider2000,alon1995}. Experiments and simulations show that the bubble coefficients $\alpha_b$ and $\theta_b$ are nearly independent of density ratio over a broad range, while the spike coefficients $\alpha_s$ and $\theta_s$ increase with $R$ \citep{alon1995,cheng2000}. In the vacuum limit $A\to1$, an RT spike approaches free fall, $\alpha_s\to1/2$, and an RM spike approaches ballistic motion, $\theta_s\to1$ \bluetext{\citep{cheng2000,alon1995}}. These asymptotic laws are stringent requirements on any physically admissible buoyancy--drag model.

Several levels of modelling are available. Algebraic similarity laws are inexpensive and capture asymptotic growth, but their coefficients must still be supplied by data or auxiliary theory, and they do not describe the full response to complex initial conditions or time-dependent forcing \citep{zhou2019}. At the other end of the spectrum, direct numerical simulation, large-eddy simulation (LES) \citep{xiao2022les,xu2026adaptive} and Reynolds-averaged Navier--Stokes (RANS) methods resolve or model spatially distributed flow fields. Variable-density RANS closures can be grouped broadly into two-equation models, such as $k$--$\varepsilon$ and $K$--$L$, and BHR-type second-moment closures that evolve quantities including the turbulent mass flux and density--specific-volume covariance. The former represent turbulent kinetic energy together with a dissipation or length scale using relatively few transport equations; the latter retain more of the variable-density coupling and can represent asymmetric transport in greater detail \citep{xie2023kl,zhao2025four,xiao2021kl,xie2021bhr,xie2021kepsilon,xie2025compressible,zhang2020}.

Such field models are more expensive than low-order descriptions and are less convenient for rapid exploration of large parameter spaces \citep{zhou2017,boffetta2017}. Buoyancy--drag ordinary differential equations occupy an intermediate level: they do not resolve the full flow, but reduce front motion to a balance among inertia, buoyancy and drag. They retain the classical similarity scalings while accepting a measured acceleration history directly as input \bluetext{\citep{hansom1990,dimonte1996,srebro2003}}. The present study concerns this class of zero-dimensional models. Depending on the initial disturbance and the mixing state, buoyancy--drag models may describe either a single mode or multimode mixing. We focus on multimode models because practical interfaces generally contain a broad range of scales.

Two complementary modelling viewpoints have emerged over several decades \citep{zhou2017}. Front-structure models attribute width growth to the large bubbles and spikes at the edges of the mixing region. Their single-mode foundation can be traced to Layzer's potential-flow bubble model \citep{layzer1955}. Youngs and co-workers extended this local force picture to multimode mixing by evolving the bubble and spike boundaries with zero-dimensional buoyancy--drag equations whose length scale grows with the layer \citep{hansom1990}. Subsequent developments include the independent cylindrical-front model of \citet{dimonte1996}, the density-ratio-dependent coefficients of \citet{cheng2000}, and the general time-dependent-acceleration formulation of \citet{srebro2003}. A second viewpoint treats the layer as a fully developed, spanwise homogeneous turbulent mixture and couples the bubble and spike widths through mean density profiles, mass conservation or approximate momentum constraints. Representative examples are the spanwise homogeneous model of \citet{dimonte2000}, the nonlinear mean-profile model of \citet{zhang2016}, and the mass-conserving extension of \citet{li2025}.

The remaining difficulty is physical closure of the model coefficients over wide density ratios and unsteady forcing. Front models retain the local dynamics, but commonly prescribe added mass, drag and the characteristic scale empirically. Statistical models couple the two sides through a mean profile and conservation, but do not supply the local front dynamics. We therefore use the local force balance to determine the equation structure and then close the two sides with global mass conservation, a mean-composition profile, similarity scalings and endpoint conditions. A second issue is state multiplicity: at a fixed density ratio, the self-similar coefficients can still depend on the initial perturbation and loading history. We represent this non-uniqueness by a state parameter $c$ that selects the internal shape of the mean-composition profile.

The model advances the low-order description in three respects. First, separate bubble and spike equations retain the opposite contributions of entrainment and added mass to the effective inertia on the two sides. Second, the bubble and spike are not forced to share inertia, buoyancy or drag coefficients; their density-ratio dependence is determined jointly from RT/RM similarity solutions, mass conservation, the mean profile and endpoint asymptotics. Third, RT spike scaling alone selects the mean-profile state, after which an explicit RT--RM closure predicts the RM spike exponent. RM spike data are not used to recalibrate either the state branch or the model coefficients.

The paper is organised as follows. Section~\ref{sec:2} reviews front-structure and statistical-mean approaches. Section~\ref{sec:3} develops the model and distinguishes state inputs, coefficient inversion and the RT--RM cross-problem closure. Section~\ref{sec:4} tests RT scaling and temporal evolution, the RM spike exponent and post-impulse evolution, variable-acceleration cases, and the Nova deceleration stage. Section~\ref{sec:5} discusses identifiability, mixing-state branches and the range of applicability. Section~\ref{sec:6} summarises the conclusions. The appendices provide the endpoint expansion and a sensitivity test for the RT--RM mapping.

\section{Development of multimode buoyancy--drag models}\label{sec:2}

Multimode buoyancy--drag models usually take the bubble or spike penetration width as their primary unknown. With $v_i\equiv dh_i/dt$ ($i=b,s$ denoting bubble and spike), a general form is
\begin{equation}
f_{i 1} ( \rho_{1} , \rho_{2} ) L_{i} S_{i} \frac{d v_{i}}{d t} = f_{i 2} ( \rho_{1} , \rho_{2} ) L_{i} S_{i} g - f_{i 3} ( \rho_{1} , \rho_{2} ) S_{i} v_{i} | v_{i} |
\label{eq:2-1}
\end{equation}
Here $L_iS_i$, $S_i$ and $L_i$ are the characteristic volume, cross-sectional area and length of the representative structure. The left-hand side of (\ref{eq:2-1}) is inertia; the first and second terms on the right are buoyancy and drag. The functions $f_{i1}$, $f_{i2}$ and $f_{i3}$ contain the density dependence of these terms. Inertia and buoyancy scale with volume, whereas drag scales with frontal area; $v_i|v_i|$ ensures the correct drag direction. Models differ primarily in how these functions are closed and in whether the bubble and spike equations evolve independently or are coupled. \bluetext{Other low-order descriptions include energy-balance models \citep{youngs2013}, bubble-competition models \citep{alon1995,shvarts1995}, and characteristic-wavelength evolution models \citep{srebro2003}.} The discussion below is restricted to Newtonian buoyancy--drag models that directly evolve mixing widths.

\bluetext{Within the Newtonian buoyancy--drag class, two complementary closure routes have emerged. Front-structure models apply the force balance to representative bubble and spike fronts, whereas statistical-mean models infer effective inertia and buoyancy from averaged mixing-layer profiles and couple the two widths through global constraints; these routes are reviewed in \S~\ref{sec:2-1} and \S~\ref{sec:2-2}, respectively.}

\subsection{Front-structure dynamics}\label{sec:2-1}

This viewpoint attributes multimode width growth to the large bubble and spike structures at the layer edges and writes buoyancy--drag equations for those representative fronts. Its theoretical origin is Layzer's 1955 single-mode potential-flow model \citep{layzer1955}. In the vacuum limit $\rho_1=0$ ($A=1$), Layzer considered a bubble rising in a cylindrical channel of diameter $D$. With bubble-tip position $h_b$, velocity $v_b=dh_b/dt$, characteristic length $L_b=D$ and frontal area $S_b$, the dimensional form corresponding to (\ref{eq:2-1}) can be written as \bluetext{\citep{layzer1955,hansom1990}}
\begin{equation}
\bluetext{\rho_{2} ( 1 + E_{b} ) L_{b} S_{b} \frac{d v_{b}}{d t} = \rho_{2} ( 1 - E_{b} ) L_{b} S_{b} g - 2 j_{1,1} \rho_{2} S_{b} v_{b} | v_{b} | .}
\label{eq:2-2}
\end{equation}
\bluetext{Here $j_{1,1}=3.8317$ is the first positive zero of the first-order Bessel function $J_1$, and $E_b=\exp(-4j_{1,1}h_b/D)$ is the finite-amplitude correction from Layzer's potential-flow solution.} The left-hand side has the form density $\times$ volume $\times$ acceleration; the terms on the right are buoyancy acting on $L_bS_b$ and drag acting on $S_b$. \bluetext{Thus $f_{b1}=\rho_2(1+E_b)$, $f_{b2}=\rho_2(1-E_b)$ and $f_{b3}=2j_{1,1}\rho_2$. In the strongly nonlinear regime $h_b/D\gg1$, $E_b\to0$ and the bubble approaches the terminal velocity $v_b\to\sqrt{gD/(2j_{1,1})}$.} Layzer's model therefore established the local balance ``inertia equals buoyancy minus drag'', but only for a single $A=1$ bubble with fixed lateral scale. \bluetext{Subsequent potential-flow formulations extended this single-mode construction to finite density ratios and broader RT/RM settings \citep{hecht1994,mikaelian1998,goncharov2002}.} Multimode mixing at finite density ratio requires evolving length scales and separate bubble and spike force balances. 

\subsubsection{Youngs' early buoyancy--drag model (1990)}\label{sec:2-1-1}
Youngs made the key step from a fixed-scale single bubble to multimode mixing. As reported by \citet{hansom1990}, the model omitted the small-amplitude exponential $E_b$, evolved separate bubble and spike boundaries, and replaced $D$ by the common dynamic scale $L_b=L_s=\max(h_b,h_s)$. In the notation of (\ref{eq:2-1}), $f_{b1}=\rho_1+C_M^b\rho_2$, $f_{b2}=\rho_2-\rho_1$ and $f_{b3}=C_D^bF_b$ for bubbles, while $f_{s1}=\rho_2+C_M^s\rho_1$, $f_{s2}=\rho_2-\rho_1$ and $f_{s3}=C_D^sF_s$ for spikes. \bluetext{Here $C_M^b$ and $C_M^s$ denote the bubble and spike added-mass coefficients, respectively, while $C_D^b$ and $C_D^s$ are the corresponding drag coefficients.} Both inertial densities were $\rho_1+\rho_2$, equivalent to $C_M^b=C_M^s=1$. A common drag coefficient $C_D^b=C_D^s=3.667$ was calibrated to the incompressible rocket-rig relation $h_b\simeq0.06Agt^2$ \bluetext{\citep{read1984,hansom1990}}. Bubble--spike asymmetry entered through the empirical drag-density factors $F_b=\rho_1+\rho_2$ and $F_s=(\rho_1+\rho_2)R^{-0.51}$. The formulation accommodates time-dependent density and acceleration and couples both fronts through their common scale. Its shared scale and constant added-mass and drag coefficients, however, leave the density-ratio-dependent front asymmetry largely in an empirical drag factor and provide only a limited treatment of impulsively driven RM mixing.

\subsubsection{\bluetext{Dimonte's independent cylindrical-front model (1996)}}\label{sec:2-1-2}

Unlike Youngs' common-scale coupling, \citet{dimonte1996} represented the two fronts as independently evolving cylindrical mixing structures and set $L_b=h_b$ and $L_s=h_s$. The bubble functions are $f_{b1}=\rho_2+\rho_1$, $f_{b2}=C_B(\rho_2-\rho_1)$ and $f_{b3}=C_D\rho_2$; their spike counterparts are $f_{s1}=\rho_1+\rho_2$, $f_{s2}=C_B(\rho_2-\rho_1)$ and $f_{s3}=C_D\rho_1$. \bluetext{Here $C_B$ and $C_D$ denote the common buoyancy and drag coefficients, respectively. This is equivalent to $C_M^b=C_M^s=1$, with $C_B=0.75$ and $C_D=5.5$ calibrated using sustained- and impulsive-acceleration measurements \citep{dimonte1996}.} The independent equations yield simple analytical scalings, but the shared constants produce a stronger density-ratio dependence of the RT coefficients and RM exponents than observed and do not recover the exact free-fall spike limit. Independent front scales alone are therefore insufficient without a density-dependent asymmetric closure.

\subsubsection{The density-ratio-dependent model of Cheng et al. (2000)}\label{sec:2-1-3}

To improve the density-ratio dependence and bubble--spike coupling, \citet{cheng2000} retained $L_b=h_b$ and $L_s=h_s$ but introduced distinct $C_D^b(A)$ and $C_D^s(A)$. Their functions are $f_{b1}=\rho_1+C_M^b\rho_2$, $f_{b2}=\rho_2-\rho_1$, $f_{b3}=C_D^b\rho_2$ and $f_{s1}=\rho_2+C_M^s\rho_1$, $f_{s2}=\rho_2-\rho_1$, $f_{s3}=C_D^s\rho_1$. The added-mass coefficients are assigned from front geometry: $C_M^b=C_M^s=1$ for cylindrical fronts connected to the parent phase and $1/2$ for detached three-dimensional structures. The RT bubble coefficient $\alpha_b$ determines $C_D^b(A)$ by inversion. The spike drag follows from a stationary-centre-of-mass approximation, $Z_{\mathrm{COM}}=\alpha_{\mathrm{COM}}Agt^2$ with $\alpha_{\mathrm{COM}}=k_{\mathrm{COM}}A^\gamma$, where $\gamma$ is fitted to experiment. \redtext{This construction brought density-ratio dependence, bubble--spike coupling and RT--RM transfer into a single equation set with fewer independently calibrated quantities, although the centre-of-mass relation remained empirical. This line of work was subsequently extended to the dynamical evolution of mixing fronts, theoretical methods for determining mixing and multiphase-flow descriptions, and later to a reassessment of the RT/RM similarity coefficients \citep{cheng2002dynamics,cheng2003methods,cheng2005multiphase,cheng2020alphatheta}.}

\subsection{Statistical-mean mixing-layer models}\label{sec:2-2}

The second viewpoint treats the bubble and spike regions as continua within a fully developed turbulent layer rather than tracking individual tips. Mean density profiles provide the inertia and buoyancy, while mass or momentum constraints couple the two penetration widths.

\subsubsection{Dimonte's spanwise homogeneous mixing-layer model (2000)}\label{sec:2-2-1}

To reduce the empiricism of front-structure models, \citet{dimonte2000} treated the bubble and spike regions as spanwise homogeneous mixed fluids. A piecewise-linear mean density profile and mass conservation give the interfacial density $\rho_I=(\rho_1h_s+\rho_2h_b)/(h_s+h_b)$ and the mean dynamic densities $\bar\rho_b=(\rho_2+\rho_I)/2$ and $\bar\rho_s=(\rho_1+\rho_I)/2$. Because $\rho_I$ depends on both widths, mass conservation couples their evolution. With $L_b=h_b$ and $L_s=h_s$, the functions are $f_{b1}=\bar\rho_b$, $f_{b2}=\rho_2-\bar\rho_b$, $f_{b3}=C_D^b\rho_2$ and $f_{s1}=\bar\rho_s$, $f_{s2}=\bar\rho_s-\rho_1$, $f_{s3}=C_D^s\rho_1$. \bluetext{The profile fixes inertia and buoyancy, leaving a single common drag constant, $C_D^b=C_D^s=C\approx2.5\pm0.6$, calibrated against linear electric motor (LEM) data \citep{dimonte2000}.} The model captures the broad density-ratio trends of RT coefficients and RM exponents, but its RM prediction depends on the initial bubble--spike width ratio and the drag constant requires some case-to-case adjustment. It nevertheless established an important connection between global mass partition and local dynamics.

\subsubsection{The nonlinear-profile and quasi-momentum model of Zhang et al. (2016)}\label{sec:2-2-2}

\citet{zhang2016} replaced Dimonte's piecewise-linear profile by nonlinear and potentially dissimilar profiles on the two sides, thereby accounting for multiple pairs of growth coefficients at a fixed density ratio. The mean density was written as $\bar\rho_i=w_i\rho_I+(1-w_i)\rho_i$, with $\rho_I$ obtained from mass conservation, and the profile asymmetry was measured by $\eta=w_s/w_b$. Only the bubble equation was evolved; \bluetext{the spike followed from the quasi-momentum conservation constraint $\bar\rho_s h_s\,\mathrm d\bar V_s/\mathrm dt=\bar\rho_b h_b\,\mathrm d\bar V_b/\mathrm dt$}, closed by the parabolic relation $\bar V_i=v_i/3$. Thus $f_{b1}=\bar\rho_b$, $f_{b2}=3(\rho_2-\bar\rho_b)$ and $f_{b3}=3C_d\rho_2$, with $C_d\approx0.83$ and $L_b=h_b$. The model explained the spread of growth coefficients and performed well under several acceleration histories. Its quasi-momentum relation is not an exact conservation law for continuously accelerated RT mixing, however, and the state-dependent free parameter $\eta$ does not enforce the spike asymptote as $A\to1$.

\subsubsection{The mass-conserving model of Li and Zhang (2025)}\label{sec:2-2-3}

\citet{li2025} replaced the quasi-momentum relation by exact mass conservation. Building on the density-ratio-invariant mean-species profile of \citet{ruan2020}, they derived an analytical mean density profile and an algebraic relation between $h_s$ and $h_b$, so that only the bubble requires a dynamical equation. Their functions are $f_{b1}=\bar\rho_b$, $f_{b2}=\rho_2-C_B^b(A)\bar\rho_b$ and $f_{b3}=3\bar\rho_b$, with $L_b=h_b$. The profile-derived $\bar\rho_b$ represents the inertia of the bubble region. Entrainment is incorporated through $C_B^b(A)=1-0.3A$, fixed by the RT asymptote near $A=0$, while the drag follows by inversion from the LEM value $\theta_b=0.25$. The model uses one coefficient set for sustained, impulsive and variable-acceleration LEM cases and for the Nova deceleration stage. It does not, however, recover the spike asymptote as $A\to1$.

\subsection{Synthesis and present approach}\label{sec:2-3}

Front-structure and statistical-mean models address different levels of the same problem. The former preserve the local force balance and provide a direct route to high-density-ratio limits, but often require empirical added mass, drag and length scales. The latter couple $h_b$ and $h_s$ through a mean profile and conservation, but their predictions depend on the profile closure and a global constraint cannot replace local front dynamics. Their strengths are therefore complementary. The present model retains separate bubble and spike force balances while using global mass conservation to couple their evolution.

\section{A buoyancy--drag model coupling local front dynamics and global conservation}\label{sec:3}

The construction proceeds from local dynamics to global closure. Separate force balances are first written for representative bubble and spike fronts. Under quasi-incompressible conditions, a mean-composition profile and mass conservation then relate the two penetration widths. Finally, RT and RM similarity solutions, symmetry as $A\to0$, and vacuum-limit asymptotics as $A\to1$ map the observable scalings onto the model coefficients. This preserves the mixing-state information carried by the mean profile while closing the local front equations through global constraints.

\subsection{Local model: bubble and spike front dynamics}\label{sec:3-1}

\begin{figure}
\centering
\includegraphics[width=0.86\textwidth]{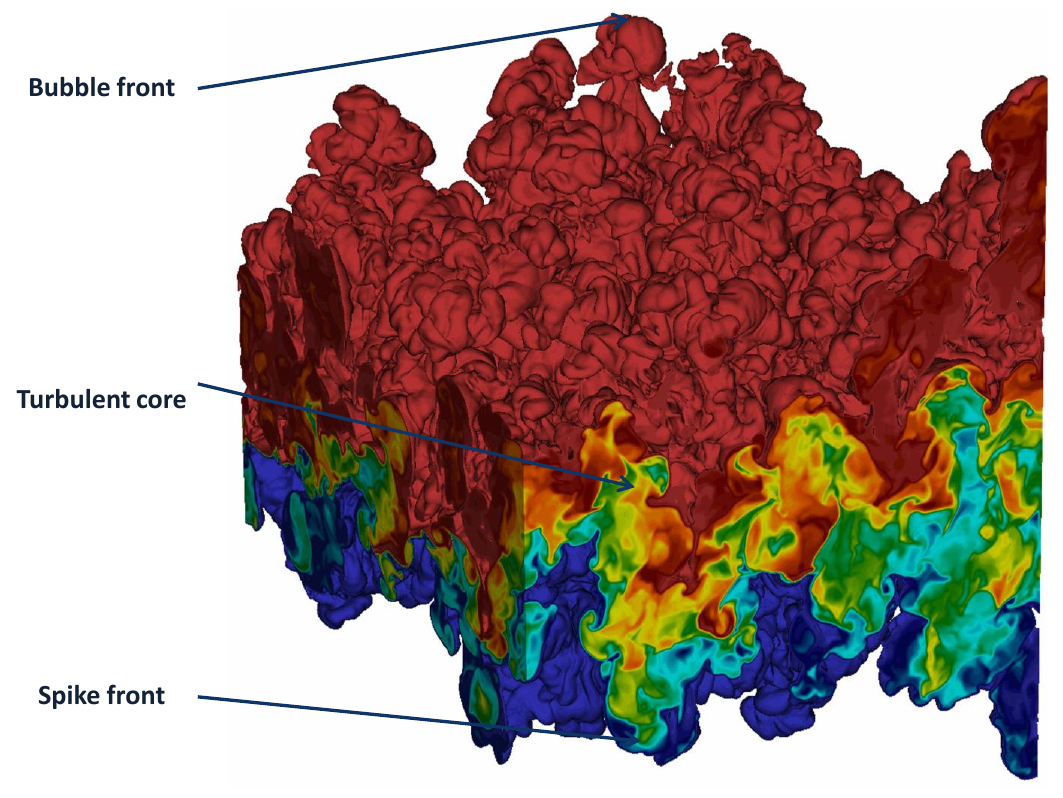}
\caption{Illustrative three-dimensional snapshot of a fully developed RT mixing layer, showing the bubble front, turbulent core and spike front used in the local model. The rendering uses a previously unpublished frame from the LES described by \citet{xiao2024localtransition}, provided by M.-J. Xiao.}\label{fig:1}
\end{figure}

Figure~\ref{fig:1} illustrates a fully developed RT layer and the front regions modelled here. Once self-similar growth is established, a turbulent core occupies the central part of the layer, while large-scale bubbles and spikes control its two edges. Because $h_b$ and $h_s$ are the displacements of these fronts, each representative structure is assigned a characteristic volume and frontal area. Newton's second law then provides separate bubble and spike buoyancy--drag equations.

Entrainment and added mass are kept distinct at the mechanical level and combined only through their net contribution to front inertia. Normalising the equation by an effective inertial volume makes the buoyancy coefficient the ratio of the body volume on which buoyancy acts to that effective volume; it therefore need not equal unity. The mean profile and global mass conservation will subsequently constrain the two widths and their coefficients.

\subsubsection{Dynamical equations for the bubble and spike fronts}\label{sec:3-1-1}

The same mechanical decomposition is used on both sides, while allowing distinct scales and coefficients to represent asymmetric evolution. We first derive the bubble equation and then obtain the spike equation by interchanging the fluid roles.

Consider a light-fluid bubble penetrating heavy fluid. Let $V_0^b$ and $S_0^b$ be its body volume and frontal area, and let $h_b$ and $v_b\equiv\dot h_b$ be its displacement and speed. Archimedean buoyancy acts on the displaced body volume, $F_B=(\rho_2-\rho_1)V_0^bg$. Drag is based on the dynamic pressure of the pure heavy fluid penetrated by the bubble, rather than on the mean density of the mixed structure: $F_D=\tfrac12C_D^{\mathrm{phys}}\rho_2S_0^bv_b^2$. The inertia contains two physically distinct contributions. Kelvin--Helmholtz shear and large eddies entrain heavy fluid into the bubble structure, while acceleration of the bubble also sets part of the surrounding heavy fluid in motion without mixing it. Thus $F_I=(\rho_1+C_E^b\rho_2)V_0^b\dot v_b+C_M^b\rho_2V_a^b\dot v_b$, where $C_E^b$ measures the entrainment contribution, $C_M^b$ is an added-mass coefficient, and $V_a^b$ is the effective volume of co-accelerated fluid. Newton's law $F_I=F_B-F_D$ gives
\begin{equation}
( \rho_{1} + C_{E}^{b} \rho_{2} ) V_{0}^{b} \dot v_{b} + C_{M}^{b} \rho_{2} V_{a}^{b} \dot v_{b} = ( \rho_{2} - \rho_{1} ) V_{0}^{b} g - \tfrac12 C_{D}^{\mathrm{phys}} \rho_{2} S_{0}^{b} v_{b}^{2}
\label{eq:3-1}
\end{equation}
Introduce the net inertia coefficient $C_I^b\equiv C_M^b+C_E^b$ and define an effective inertial volume $\widetilde V^b$ such that the left-hand side of (\ref{eq:3-1}) is identically $(\rho_1+C_I^b\rho_2)\widetilde V^b\dot v_b$. With $r(A)\equiv V_a^b/V_0^b$, this definition gives $\widetilde V^b=V_0^b[1+C_E^bR+C_M^brR]/[1+C_I^bR]$. This is an exact rewriting, not an empirical approximation. The physical volumes are $V_0^b$ and $V_a^b$; $\widetilde V^b$ is the equivalent volume defined by their combined inertia. Normalisation by $\widetilde V^b$ yields
\begin{equation}
( \rho_{1} + C_{I}^{b} \rho_{2} ) \dot v_{b} = C_{B}^{b} ( \rho_{2} - \rho_{1} ) g - C_{D}^{b} \rho_{2} v_{b}^{2} / h_{b}
\label{eq:3-2}
\end{equation}
where $C_D^b\equiv\tfrac12C_D^{\mathrm{phys}}S_0^bh_b/\widetilde V^b$ incorporates frontal area, length and effective volume. The coefficient entering the drag term is therefore $C_D^b\rho_2$. The buoyancy coefficient is
\begin{equation}
\redtext{C_{B}^{b} \equiv V_{0}^{b} / \widetilde{V}^{\,b} = ( 1 + C_{I}^{b} R ) / ( 1 + C_{E}^{b} R + C_{M}^{b} r R )}
\label{eq:3-3}
\end{equation}
$C_B^b$ is the ratio of the body volume on which buoyancy acts to the effective inertial volume. It equals unity only when $r(A)=V_a^b/V_0^b=1$ and may otherwise vary with density ratio. This is a lumped description of a self-similar front: entrained mass flux and unresolved momentum flux are incorporated into effective coefficients. Accordingly, $C_E$ represents the net entrainment contribution to low-order inertia, not a local entrainment rate. Available width data identify only the combination $C_I^b$, not $C_E^b$ and $C_M^b$ separately.

For a spike, the body and penetrated-fluid densities are interchanged and entrainment has the opposite sign in the net inertia:
\begin{equation}
\redtext{( \rho_{2} + C_{I}^{s} \rho_{1} ) \dot v_{s} = C_{B}^{s} ( \rho_{2} - \rho_{1} ) g - C_{D}^{s} \rho_{1} v_{s}^{2} / h_{s}}
\label{eq:3-4}
\end{equation}
The spike body is heavy fluid and penetrates light fluid, so the drag density is $\rho_1$. Entrainment of light fluid reduces the spike inertia, giving $C_I^s=C_M^s-C_E^s$, in contrast to $C_I^b=C_M^b+C_E^b$. Its buoyancy coefficient is defined by the analogous volume ratio,
\begin{equation}
\redtext{C_{B}^{s} \equiv V_{0}^{s} / \widetilde{V}^{\,s} = ( R + C_{I}^{s} ) / ( R - C_{E}^{s} + C_{M}^{s} V_{a}^{s} / V_{0}^{s} )}
\label{eq:3-5}
\end{equation}
Again, $C_B^s=1$ only if $V_a^s/V_0^s=1$; the two volume ratios are not assumed equal. The subsequent closure uses only the bubble value $r(0)$ at the symmetric endpoint. \redtext{The equation is compatible with the vacuum limit: if $C_I^s\to0$, $C_B^s\to1$ and $C_D^s\rho_1\to0$, then (\ref{eq:3-4}) reduces to $\dot v_s=g$ for RT and $\dot v_s=0$ for post-impulse RM, thereby naturally recovering the RT free-fall limit and the constant-velocity inertial limit of post-impulse RM.}

In the notation of (\ref{eq:2-1}), the bubble functions are $f_{b1}=\rho_1+C_I^b\rho_2$, $f_{b2}=C_B^b(\rho_2-\rho_1)$ and $f_{b3}=C_D^b\rho_2$; the spike functions are $f_{s1}=\rho_2+C_I^s\rho_1$, $f_{s2}=C_B^s(\rho_2-\rho_1)$ and $f_{s3}=C_D^s\rho_1$. The equations are homogeneous of degree one in density, so only $R$, or equivalently $A$, is independent. Henceforth we normalise by $\rho_1+\rho_2$ and use $\rho_1=(1-A)/2$ and $\rho_2=(1+A)/2$.

The local model thus contains six effective coefficients, one inertia, buoyancy and drag coefficient for each front. Retaining all six allows distinct density-ratio dependence; it does not make them independent fitting parameters. Section~\ref{sec:3-2} closes them from bubble scalings, symmetric-endpoint geometry, a mean-profile state, similarity relations, mass conservation and endpoint conditions.

\subsection{Global closure: similarity scaling, mass conservation and the mean profile}\label{sec:3-2}

Although the advance of the bubble and spike fronts directly sets the two penetration widths, the fronts bound a single continuous mixing layer and cannot evolve independently. They must jointly satisfy mass conservation across the layer. Under the quasi-incompressible approximation, the mean-composition profile also fixes the mean-density distribution and thereby converts global mass conservation into a quantitative relation between the bubble and spike widths. We therefore treat the entire layer, including its fronts and turbulent core, as the statistical control volume, and combine similarity scalings with endpoint constraints to close the six effective coefficients in the local dynamical equations.

The closure proceeds in four stages. First, the classical quadratic RT solution and power-law RM solution are substituted into the front equations to invert the model coefficients in terms of similarity parameters; the bubble scalings are supplied as experimental inputs. Second, an idealised geometric estimate at the symmetric endpoint fixes the bubble inertia coefficient, while symmetry and vacuum-limit constraints yield a minimum-order even representation for its spike counterpart. Third, mass conservation and the mean-composition profile determine the RT spike growth coefficient from the prescribed bubble coefficient. Finally, consistency at the common RT/RM endpoints and first-order matching at the vacuum limit map the RT spike scaling to the RM spike exponent. Once the bubble scalings and the mixing-state parameter are specified, all six effective coefficients follow from this common set of physical constraints.

\subsubsection{Relations between similarity scalings and model coefficients}\label{sec:3-2-1}

For classical RT mixing under constant acceleration, the buoyancy--drag equations admit the quadratic similarity solution $h=\alpha Agt^2$. After an impulsive acceleration, $g=0$ and RM mixing admits the power-law solution $h\propto t^\theta$. Substitution of these solutions into the bubble equation~\eqref{eq:3-2} and spike equation~\eqref{eq:3-4} gives the buoyancy and drag coefficients as functions of the Atwood number (or, equivalently, $R$), the similarity parameters and the corresponding inertia coefficient:
\begin{equation}
C_{B}^{b} ( A ) = [ 1 + C_{I}^{b} ( A ) R ] 2 \alpha_{b} ( 2 - \theta_{b} ) / ( ( R + 1 ) \theta_{b} )
\label{eq:3-6}
\end{equation}
\begin{equation}
C_{D}^{b} ( A ) = C_{B}^{b} ( A ) ( R + 1 ) ( 1 - \theta_{b} ) / ( 2 \alpha_{b} ( 2 - \theta_{b} ) R )
\label{eq:3-7}
\end{equation}
\begin{equation}
C_{B}^{s} ( A ) = [ C_{I}^{s} ( A ) + R ] 2 \alpha_{s} ( 2 - \theta_{s} ) / ( ( R + 1 ) \theta_{s} )
\label{eq:3-8}
\end{equation}
\begin{equation}
C_{D}^{s} ( A ) = C_{B}^{s} ( A ) ( R + 1 ) ( 1 - \theta_{s} ) / ( 2 \alpha_{s} ( 2 - \theta_{s} ) )
\label{eq:3-9}
\end{equation}
Equations~\eqref{eq:3-6} and~\eqref{eq:3-8} give the bubble and spike buoyancy coefficients, respectively, while equations~\eqref{eq:3-7} and~\eqref{eq:3-9} give the corresponding drag coefficients. Thus, both coefficients are expressed through the similarity parameters and inertia. The bubble exponents are approximately constant over a broad range of density ratios and can be fixed independently from experiment or simulation; once $C_I^b$ is known, $C_B^b$ and $C_D^b$ follow. By contrast, $\alpha_s$ and $\theta_s$ vary strongly with density ratio and are closed below.

\subsubsection{Closure of the inertia coefficients}\label{sec:3-2-2}

The effective inertia coefficient measures the combined entrainment and added-mass contribution relative to the inertia of the coherent front. Once density has been written explicitly in the inertia function, this coefficient primarily represents dimensionless front geometry, entrainment and the ratio of displaced-fluid volume to coherent-front volume. Its Atwood-number dependence therefore originates in changes of front geometry and the surrounding flow. In the fully developed similarity regime, the bubble front generally retains a robust blunt-cap morphology whose dimensionless shape, frontal scale and displaced-volume ratio vary only weakly with density ratio. We accordingly set $C_I^b(A)=C_I^b(0)$. This does not make the physical bubble inertia independent of $A$: $f_{b1}=\rho_1+C_I^b\rho_2$ still depends explicitly on $A$ through $\rho_1=(1-A)/2$ and $\rho_2=(1+A)/2$. The approximation concerns only the dimensionless geometric and entrainment/added-mass proportions after the density dependence has been separated.

The reference value of the bubble inertia coefficient is fixed jointly by the similarity relation and a geometric estimate at $A=0$. At this endpoint the fluid densities are equal, and interchanging the light- and heavy-fluid labels must leave the bubble and spike statistics unchanged. Added mass retains the same inertial role under this interchange, whereas entrainment enters the two effective inertias with opposite signs. The minimum symmetric-endpoint closure is therefore to neglect its leading-order net correction to the lumped inertia at $A=0$, i.e. $C_E^b(0)=0$. The condition $C_E^b(0)=0$ does not mean that local entrainment or fluid mixing disappears; it means only that its net correction to the lumped inertia is neglected in the present low-order front equation at the symmetric endpoint. Since $C_I^b=C_M^b+C_E^b$, it follows that $C_I^b(0)=C_M^b(0)$. With $R=1$, equation~\eqref{eq:3-3} then becomes $C_B^b(0)=[1+C_I^b(0)]/[1+C_I^b(0)r(0)]$. Equating this expression to the similarity inversion~\eqref{eq:3-6} yields
\begin{equation}
C_{I}^{b} ( 0 ) = [ \theta_{b} / ( \alpha_{b} ( 2 - \theta_{b} ) ) - 1 ] / r ( 0 )
\label{eq:3-10}
\end{equation}
Thus $C_I^b(0)$ is fixed by $r(0)$, $\alpha_b$ and $\theta_b$. We estimate $r(0)$ using a standard idealised geometry: a hemispherical bubble cap of diameter $D$ displaces a cylindrical volume of fluid of diameter and height $D$. The coherent-front volume is then $\pi D^3/12$ and the displaced volume is $\pi D^3/4$, giving $r(0)=3$.

The spike geometry is considerably more sensitive to density ratio. As $A$ increases, the spike becomes increasingly slender, while the density of the light fluid ahead of it falls and the entrainment and added-mass contributions weaken. A constant spike inertia coefficient is therefore inappropriate. We construct its $A$ dependence from three conditions: (i) symmetry requires $C_I^s(0)=C_I^b(0)$; (ii) at the vacuum limit, the light fluid ahead of the spike disappears and hence $C_I^s(1)=0$; and (iii) fluid-interchange symmetry excludes a linear bias near $A=0$, so $\left.dC_I^s/dA\right|_{A=0}=0$. These conditions do not uniquely fix the interior function. To avoid an additional empirical parameter, we choose the lowest-order monotone even polynomial satisfying all three:
\begin{equation}
C_{I}^{s} ( A ) = C_{I}^{b} ( 0 ) ( 1 - A^{2} )
\label{eq:3-11}
\end{equation}
This expression joins continuously to the bubble coefficient at $A=0$, decreases monotonically with $A$, and vanishes at the vacuum limit.

The two inertia coefficients are now closed. Equations~\eqref{eq:3-6}--\eqref{eq:3-9} reduce the remaining buoyancy and drag closure to determining the four similarity parameters $\alpha_b$, $\theta_b$, $\alpha_s$ and $\theta_s$. The bubble values are supplied by experiment; the next two subsections determine the spike values.

\subsubsection{The quadratic spike growth coefficient $\alpha_s(A)$}\label{sec:3-2-3}

Unlike the approximately density-ratio-independent bubble coefficient $\alpha_b$, the spike coefficient $\alpha_s$ varies strongly as the spike morphology changes with $A$. Capturing this variation is a central difficulty for buoyancy--drag models. Three robust constraints are available: $\alpha_s=\alpha_b$ at the symmetric endpoint, $\alpha_s$ increases monotonically with $A$, and $\alpha_s\to1/2$ at the free-fall limit. We therefore define the normalised transition
\begin{equation}
\Theta_{\alpha} ( A ) \equiv ( \alpha_{s} ( A ) - \alpha_{b} ) / ( \alpha_{s} ( 1 ) - \alpha_{b} )
\label{eq:3-12}
\end{equation}
which maps the symmetric state at $A=0$ to zero and the free-fall endpoint at $A=1$ to unity. It follows that
\begin{equation}
\alpha_{s} ( A ) = \alpha_{b} + ( 1 / 2 - \alpha_{b} ) \Theta_{\alpha} ( A )
\label{eq:3-13}
\end{equation}
The problem is thereby reduced to determining $\Theta_\alpha$.

For classical self-similar RT growth, $h_b=\alpha_bAgt^2$ and $h_s=\alpha_sAgt^2$, so $\chi\equiv h_s/h_b=\alpha_s/\alpha_b$ and
\begin{equation}
\Theta_{\alpha} ( A ) = \alpha_{b} [ \chi ( A ) - 1 ] / ( 1 / 2 - \alpha_{b} )
\label{eq:3-14}
\end{equation}
It remains to determine $\chi(A)$. Let $x_b$ and $x_s$ denote the bubble and spike fronts, respectively, and define $X=(x-x_b)/(x_s-x_b)$, such that $X=0$ and $1$ correspond to the two fronts. Conservation of mass per unit cross-sectional area gives
\begin{equation}
\redtext{\rho_{1} h_{s} + \rho_{2} h_{b} = \int_{x_{b}}^{x_{s}} \overline{\rho}(x)\,\mathrm dx}
\label{eq:3-15}
\end{equation}
where the mean density $\overline\rho(x)$ and mean volume fraction $\overline\Phi(X)$ are related by
\begin{equation}
\redtext{\overline{\rho}(x) = \rho_{1} + ( \rho_{2} - \rho_{1} )\overline{\Phi}(X)}
\label{eq:3-16}
\end{equation}
Substituting equation~\eqref{eq:3-16} into~\eqref{eq:3-15} and nondimensionalising yields
\begin{equation}
\redtext{\chi(A) = [1-\Pi(A)]/\Pi(A), \qquad \Pi(A) = \int_{0}^{1}\overline{\Phi}(X)\,\mathrm dX}
\label{eq:3-17}
\end{equation}
Thus the mean-volume-fraction profile fixes $\chi$ through~\eqref{eq:3-17}, and hence $\Theta_\alpha$ through~\eqref{eq:3-14}.

Following the composite-profile construction of Ruan et al.\ \citep{ruan2020}, the mean mass fraction $\overline Y$ and mean volume fraction $\overline\Phi$ are related, under the present composition averaging and incompressible pure-component densities, by
\begin{equation}
\redtext{\overline{\Phi} = \overline{Y}/[\overline{Y}+(1-\overline{Y})R]}
\label{eq:3-18}
\end{equation}
Directly prescribing either profile at arbitrary $A$ is difficult. Ruan et al.\ observed that, as $A\to0$, the mean mass- and volume-fraction profiles coincide with the analytic profile
\begin{equation}
C ( X ) = 2 ( X - 1 / 2 )^{3} - 3 / 2 ( X - 1 / 2 ) + 1 / 2
\label{eq:3-19}
\end{equation}
As $A$ increases, the two profiles separate into a hysteresis-like loop. A density-ratio-invariant composite can be constructed by weighting $\overline Y$ and $\overline\Phi$ with $F(X,A)$:
\begin{equation}
\redtext{F\overline{Y}+(1-F)\overline{\Phi}=C(X)}
\label{eq:3-20}
\end{equation}
A recent extension gives the weighting function
\begin{equation}
F ( A , X , c ) = [ 1 / 2 - A / 2 + A C ( X ) ( 1 + \beta X ( X - 1 ) ) ] [ 1 + c A ( A - 1 ) X ( X - 1 ) ]
\label{eq:3-21}
\end{equation}
\redtext{Equation~\eqref{eq:3-21} is part of a mean-composition-profile family recently developed by Zhang et al.\ as an extension of the composite construction of Ruan et al.\ \citep{ruan2020}; the full construction will be published elsewhere.} Given $A$, $\beta$ and $c$, equations~\eqref{eq:3-18}--\eqref{eq:3-21} determine $\overline\Phi(X)$ over $0\leq X\leq1$, from which equation~\eqref{eq:3-17} gives $\Pi(A)$ and $\chi(A)$. The parameter $\beta$ is fixed by the free-fall endpoint. Since $\alpha_s(1)=1/2$, one has $\chi(1)=1/(2\alpha_b)$ and $\Pi(1)=\alpha_b/(\alpha_b+1/2)$. Hence $\beta$ solves the scalar equation $G(\beta;\alpha_b)\equiv\int_0^1\overline\Phi(X;A=1,\beta)\,\mathrm dX-\alpha_b/(\alpha_b+1/2)=0$, evaluated by numerical quadrature and one-dimensional root finding for the prescribed $\alpha_b$. The parameter $c$ leaves the endpoints $A=0,1$ and $X=0,1$ unchanged but controls the interior profile shape and thereby the partition of the total width between bubbles and spikes. It is therefore a coordinate of the mixing state at a given density ratio. Its value is selected globally from density-ratio scalings measured in experiments or high-fidelity simulations, rather than fitted separately to each time history.

Thus, for given $A$ and $c$, equations~\eqref{eq:3-17}--\eqref{eq:3-21} determine $\chi(A)$, and equations~\eqref{eq:3-13}--\eqref{eq:3-14} then give $\Theta_\alpha(A)$ and $\alpha_s(A)$.

\subsubsection{The spike power-law exponent $\theta_s(A)$}\label{sec:3-2-4}

The RM spike exponent and the RT spike coefficient both reflect bubble--spike asymmetry, although they arise under different driving mechanisms. To relate their scalings, define
\begin{equation}
\Theta_{\theta} ( A ) \equiv ( \theta_{s} ( A ) - \theta_{b} ) / ( \theta_{s} ( 1 ) - \theta_{b} )
\label{eq:3-22}
\end{equation}
where $\theta_s(1)=1$. The function $\Theta_\theta$ connects the symmetric state at $A=0$ to the ballistic endpoint at $A=1$, so
\begin{equation}
\theta_{s} ( A ) = \theta_{b} + ( 1 - \theta_{b} ) \Theta_{\theta} ( A )
\label{eq:3-23}
\end{equation}
\redtext{We use the power mapping}
\begin{equation}
\Theta_{\theta} = \Theta_{\alpha}^{p ( A )}
\label{eq:3-24}
\end{equation}
\redtext{as an endpoint-constrained RT--RM cross-problem closure. It maps the asymmetric state encoded by the RT mean profile to $\theta_s(A)$ without using RM spike data for calibration.}

At the symmetric endpoint, $p(0)=1$. Expressed in terms of a signed Atwood number, fluid-interchange symmetry further requires $p(A)$ to be even near zero, hence $p'(0)=0$. Writing $p(1)=p_1$, the lowest-order monotone even polynomial satisfying these conditions is
\begin{equation}
p ( A ) = 1 - ( 1 - p_{1} ) A^{2}
\label{eq:3-25}
\end{equation}
\redtext{The remaining endpoint value $p_1$ is fixed by matching the first-order vacuum-limit behaviour of the spike closure.} At the vacuum limit no fluid remains ahead of the spike to be entrained or displaced, requiring $C_I^s\to0$ and $C_B^s\to1$. Let $\epsilon=1-A\to0$ and $\lambda\equiv\Theta_\alpha'(1^-)>0$. Substitution of the first-order expansions of equations~\eqref{eq:3-13}, \eqref{eq:3-23} and~\eqref{eq:3-24} into~\eqref{eq:3-8}, together with $C_B^s=1+O(\epsilon^2)$, gives
\begin{equation}
p_{1} = \frac{( \frac{1}{2} - \alpha_{b} ) \lambda + \frac{1}{4}}{( 1 - \theta_{b} ) \lambda}
\label{eq:3-26}
\end{equation}
Equation~\eqref{eq:3-9} then gives
\begin{equation}
\lim_{A\to1} C_D^s=(1-2\alpha_b)\lambda+\frac12
\label{eq:3-27}
\end{equation}
Appendix~\ref{app:asymptotic} gives the full algebra. This local expansion is used only to determine $p_1$ and the coefficient limits at the vacuum endpoint; all finite-$A$ curves are evaluated from the complete numerical profile. Although $C_D^s$ itself is large in this limit, the physical drag combination $C_D^s\rho_1$ vanishes. Appendix~\ref{app:mapping-sensitivity} tests the interior sensitivity using a quartic alternative with the same endpoint constraints.

\subsubsection{Density-ratio dependence of the complete coefficient set}\label{sec:3-2-5}

The complete six-coefficient closure is now specified. With the geometric input fixed at $r(0)=3$, only three state inputs are required: the bubble RT coefficient $\alpha_b$, the bubble RM exponent $\theta_b$, and the mean-profile parameter $c$. All intermediate quantities and model coefficients then follow sequentially. Table~\ref{tab:closure_roles} distinguishes prescribed inputs from quantities derived by the closure and those reserved for validation.

{\color{black}
\begin{table}
\color{black}
\centering
\small
\caption{Distinction among prescribed inputs, derived quantities and validation targets in the present closure.}
\label{tab:closure_roles}
\begin{tabular}{p{0.14\textwidth}p{0.17\textwidth}p{0.31\textwidth}p{0.27\textwidth}}
Category & Quantity & Value or relation & Role of data \\
\midrule
State input & $\alpha_b,\theta_b$ & LEM bubble scalings $0.05,0.25$ & Prescribed \\
State input & $c$ & $c=4$ from the RT spike branch $R^{0.33}$ & One global calibration \\
Geometric input & $r(0)$ & Hemispherical-cap/cylinder estimate, $r(0)=3$ & Symmetric endpoint \\
Derived & $\beta,\lambda,p_1$ & Profile and vacuum-limit constraints & Not fitted independently \\
Derived & $C_I^{b,s},C_B^{b,s},C_D^{b,s}$ & Equations~\eqref{eq:3-6}--\eqref{eq:3-11} & Joint six-coefficient inversion \\
Validation & $\theta_s$ & Equations~\eqref{eq:3-22}--\eqref{eq:3-26} & No recalibration to RM spike data \\
\end{tabular}
\end{table}
}

The distinction in table~\ref{tab:closure_roles} also defines the validation protocol. The quantities $\alpha_b$, $\theta_b$ and $c$ are prescribed for the mixing state of interest, with $c$ calibrated only once across density ratio. The parameters $\beta$, $\lambda$ and $p_1$, together with all six effective coefficients, are derived and are not subsequently adjusted. The RM spike exponent $\theta_s$ is excluded from calibration and tested as a cross-problem prediction. For a fixed initial perturbation and experimental configuration, $\alpha_b$ and $\theta_b$ are approximately constant across density ratio, but their plateau values change with initial and experimental conditions and are not universal. Youngs' high-resolution simulations give $\alpha_b\approx0.025\text{--}0.03$ for RT mixing generated by short-wave mode coupling and about $0.045\text{--}0.12$ when long waves are included \bluetext{\citep{youngs2013}}. AWE rocket-rig experiments give $\alpha_b=0.050\text{--}0.077$, with a recommended value near $0.06$ \bluetext{\citep{youngs1989,youngs2013}}, whereas the LEM experiments give $\alpha_b\approx0.05$ and $\theta_b\approx0.25$ \bluetext{\citep{dimonteschneider2000}}. The degree of asymmetry also varies: the LEM data approximately follow $\alpha_s/\alpha_b\sim R^{0.33}$ \citep{dimonteschneider2000}, while the Youngs and Kucherenko data lie on weaker branches, approximately $R^{0.2}\text{--}R^{0.231}$ \bluetext{\citep{youngs2013,kucherenko1991}}.

We refer to this non-uniqueness of similarity parameters and mean profiles at fixed density ratio as mixing-state multiplicity. The parameters $\alpha_b$ and $\theta_b$ set the overall bubble growth, while $c$ characterises the mean-profile shape and bubble--spike asymmetry. For example, $c=4$ represents the LEM $R^{0.33}$ branch, whereas $c=-4$ produces the weaker asymmetry observed by Youngs and Kucherenko \citep{youngs1991,kucherenko1991}. Such multiplicity is closely linked to the spectrum, amplitude and long-wave content of the initial perturbation \citep{ni2023spike,zhou2018}, and can also be influenced by miscibility, surface tension, finite Reynolds number, geometric confinement and loading history \bluetext{\citep{ramaprabhu2004,roberts2016}}. Differences in width definition, diagnostic threshold and the interval identified as self-similar add further scatter. No quantitative map from these factors to $\alpha_b$, $\theta_b$ and $c$ is yet available, so the present model treats them as state inputs. Establishing that map is an important next step.

\begin{figure}
\centering
\includegraphics[width=\textwidth]{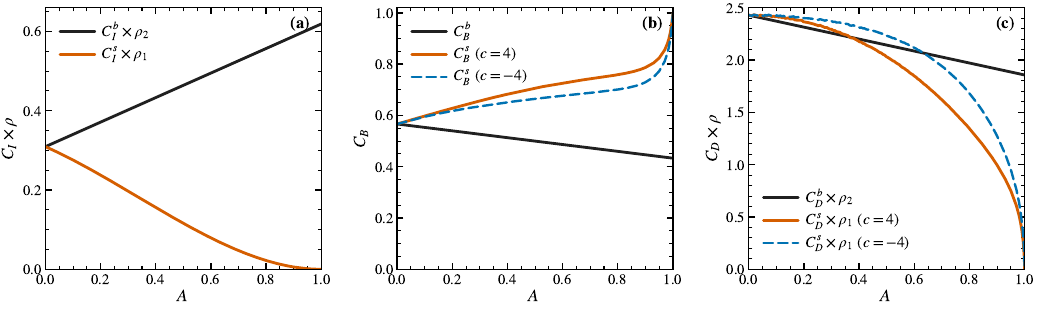}
\caption{Model coefficients and their physical combinations versus Atwood number $A$: (a) inertia contributions $C_I^b\rho_2$ and $C_I^s\rho_1$; (b) buoyancy coefficients $C_B^b$ and $C_B^s$; (c) effective drag coefficients $C_D^b\rho_2$ and $C_D^s\rho_1$. Black denotes the bubble side, orange the spike side for $c=4$, and the blue dashed curves the spike side for $c=-4$.}\label{fig:2}
\end{figure}

Because the primary validation uses the LEM experiments, we set $\alpha_b=0.05$, $\theta_b=0.25$ and $c=4$ throughout. Together with $r(0)=3$, these values give $\beta=0.113$, $C_I^b=13/21$ and $\lambda\approx250$. The vacuum-limit relation in \S~\ref{sec:3-2-4} then gives $p_1\approx0.6013$, so $p(A)=1-0.3987A^2$, and $C_D^s(1)\approx225.5$. The bubble inertia coefficient is the constant $13/21$, the spike inertia coefficient follows from~\eqref{eq:3-11}, and the buoyancy and drag coefficients follow from~\eqref{eq:3-6}--\eqref{eq:3-9}. Figure~\ref{fig:2} shows both the coefficients and the density-weighted combinations that enter the equations of motion.

The bubble buoyancy coefficient decreases monotonically from $C_B^b=17/30$ at $A=0$ to $13/30$ as $A\to1$, whereas $C_B^s$ increases from $17/30$ to unity. At the vacuum limit the spike effective volume reduces to its coherent volume, so $C_B^s\to1$; heavy fluid remains ahead of the bubble, leaving a finite added inertia and hence a finite $C_B^b$. The inertia term contains the density-weighted combination $C_I\rho_{\mathrm{ahead}}$, not $C_I$ alone. Thus $C_I^b\rho_2$ rises from $13/42$ to $13/21$, while $C_I^s\rho_1$ decreases monotonically from $13/42$ to zero. Both remain non-negative. Similarly, $C_D^b$ decreases from $34/7$ to $13/7$ and $C_D^b\rho_2$ from $17/7$ to $13/7$. Although $C_D^s$ rises from $34/7$ to about $225.5$, its physical combination $C_D^s\rho_1$ decreases from $17/7$ to zero. The spike drag therefore vanishes with the light fluid ahead of it despite the large isolated coefficient.

In summary, the density-weighted inertia and drag combinations remain non-negative for all density ratios, recover bubble--spike symmetry at $A=0$, and remove the spike added inertia and drag at $A\to1$. The combination of a fixed geometric input, similarity relations, endpoint constraints, mass conservation and the mean-composition profile thus produces smooth, asymptotically consistent coefficient evolution between the endpoints. Unlike classical closures based on constant or shared coefficients, the present model retains and physically constrains a distinct density-ratio dependence on each side.

\section{Validation, cross-prediction and application assessment}\label{sec:4}

Section~\ref{sec:3} established the equations and coefficient closure, and \S~\ref{sec:3-2-5} selected the mixing-state inputs from LEM similarity data. Using this single model and coefficient set, we now examine classical RT scaling and constant-acceleration evolution, RM spike scaling inferred from the RT closure and subsequent post-impulse evolution, transfer to variable-acceleration histories, and application to the deceleration stage of the Nova laser experiments. These tests form a progressive chain from RT closure consistency through RM cross-prediction to loading-history transfer and an external, cross-facility assessment.

\subsection{Validation data, benchmark models and numerical protocol}\label{sec:4-1}

\subsubsection{Validation data and test cases}\label{sec:4-1-1}

\redtext{The principal validation data are from the LEM experiments \citep{dimonteschneider2000}.} These measurements provide bubble, spike and total widths under constant, impulsive, increasing, decreasing and oscillatory accelerations over density ratios from approximately $1.5$ to $50$, together with the Atwood-number dependence of the RT coefficients and RM exponents. They therefore test both density-ratio dependence and unsteady response. Data from the deceleration stage of the Nova laser experiments \citep{remington1995} provide an additional test of whether a closure fixed from LEM transfers to a different facility. LEM supplies the main scaling and time-evolution benchmark; Nova serves as a stage-specific external assessment.

\subsubsection{Benchmark models and comparison protocol}\label{sec:4-1-2}

For quantitative comparisons of time evolution, we use the front-structure model of Cheng et al.\ \citep{cheng2000} and the spanwise-homogeneous mixing-layer model of Dimonte \citep{dimonte2000}, representing the local-front and statistical-mean approaches reviewed in \S~\ref{sec:2}. All three models start from the same experimental state and none is retuned case by case. The present coefficients are fixed as in \S~\ref{sec:3}. The Cheng model uses its published relations with $\alpha_b=0.05$, $C_M^b=C_M^s=1$ (the original $k=1$) and $\gamma=10$. The Dimonte model uses $C=2.0$ for every LEM case. This is not an arbitrary choice from the lower edge of an empirical interval: it is the calibration used in Dimonte's RT scaling comparison to recover the symmetric-endpoint value $\alpha_b=0.05$. The reported aggregate $C=2.5\pm0.6$ describes values adjusted across different loading histories and cases. We retain the published RT reference value $C=2.0$ to enforce a single cross-case comparison. Table~\ref{tab:model_structure} compares the dynamical structure, coupling of the two sides and closure strategy, and also lists the Li--Zhang mass-conserving model \citep{li2025}, from which the present formulation directly develops.

{\color{black}
\begin{table}
\color{black}
\centering
\small
\setlength{\tabcolsep}{3.5pt}
\caption{Dynamical structure and closure of the low-order models considered. The Li--Zhang model identifies the closest mass-conserving route to the present work but is not included in the three-model error ranking in figures~\ref{fig:5}, \ref{fig:8} and~\ref{fig:10}.}
\label{tab:model_structure}
\begin{tabular}{p{0.15\textwidth}p{0.20\textwidth}p{0.22\textwidth}p{0.34\textwidth}}
Model & Front dynamics & Coupling of two sides & Main input or closure feature \\
\midrule
Cheng & Separate bubble and spike equations & Shared reduced coefficient relation & Drag inferred from RT bubble scaling; the same relation gives RM exponents \\
Dimonte & Separate bubble and spike equations & Piecewise-linear mean-density profile & Common empirical drag coefficient $C$ \\
Li--Zhang & Bubble-side dynamics & Mean-composition profile and exact mass conservation & Profile determines the bubble--spike width partition \citep{li2025} \\
Present & Separate bubble and spike equations & Mean-composition profile and mass conservation & Six effective coefficients jointly closed by bubble scalings, profile state and endpoints \\
\end{tabular}
\end{table}
}

As table~\ref{tab:model_structure} shows, Li--Zhang shares the mean-profile and mass-conservation basis of the present model but evolves only the bubble side; the spike width is obtained algebraically from the profile constraint. Here both fronts retain local dynamical equations and their coefficients are subjected to the same scaling and endpoint constraints. Li--Zhang is therefore included to establish the lineage and extension of the mass-conserving approach, rather than as an independent two-front closure. The error rankings use Cheng and Dimonte as independent representatives of front-structure and statistical-mean modelling, respectively.

\subsubsection{Acceleration histories, initial states and numerical setup}\label{sec:4-1-3}

The LEM histories differ in amplitude, duration and waveform, so equal physical times do not represent equal accumulated forcing. Following Dimonte and Schneider \citep{dimonteschneider2000}, we compare them using the reference displacement obtained by twice integrating the measured acceleration,
\begin{equation}
Z(t)=\int_{0}^{t}\int_{0}^{t'}g(\tau)\,\mathrm{d}\tau\,\mathrm{d}t'.
\label{eq:4-1}
\end{equation}
which is the displacement of a drag-free particle initially at rest under the prescribed history, not the mixing width. Using $Z(t)$ as the abscissa measures loading progress by its accumulated kinematic effect and places constant, impulsive and variable accelerations on a common displacement scale.

\begin{figure}
\centering
\includegraphics[width=\textwidth]{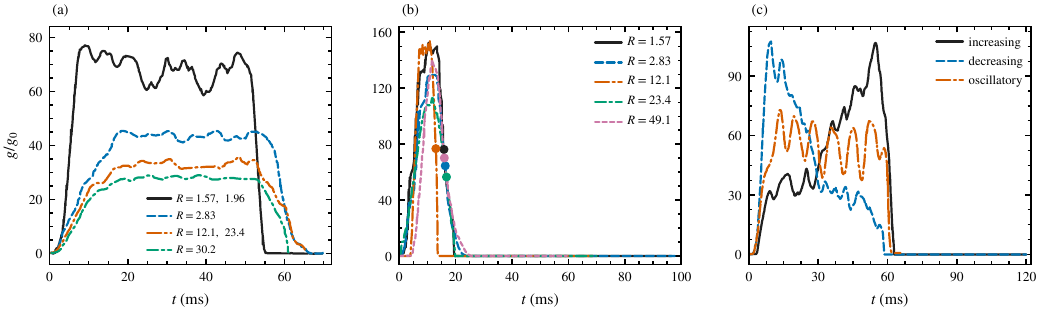}
\caption{LEM acceleration histories: (a) classical RT cases; (b) classical RM cases, with filled circles marking the hand-off time $t_*$ at 50\% of peak acceleration on the falling flank; (c) RT cases with increasing, decreasing and oscillatory acceleration. Here $g_0$ is gravitational acceleration.}
\label{fig:4}
\end{figure}

Figure~\ref{fig:4} collects the LEM acceleration histories. The classical RT cases use four distinct curves: $R=1.57$ and $1.96$ share one history, as do $R=12.1$ and $23.4$. The RM set contains five impulses, and all three variable-acceleration cases have $R=1.57$. Although $Z$ provides a common abscissa, RT and RM require different initialisation because their forcing mechanisms differ.

For classical and variable-acceleration RT, integration begins with the experimental loading and uses the full measured $g(t)$. The initial front velocities are unresolved and are set to $v_b(0)=v_s(0)=0$, consistent with the rest state used in $Z(t)$. Exactly zero widths would make the drag terms singular, so we use the finite near-zero perturbation $h_b(0)=10^{-3}\ \mathrm{cm}$, far below the widths in the comparison interval. The spike width is $h_s(0)=[\alpha_s(A)/\alpha_b]h_b(0)$, with $\alpha_s(A)$ taken from the LEM branch fixed in \S~\ref{sec:3-2-5}. This makes the initial partition compatible with the asymptotic state without introducing a second source of asymmetry. All models use these same initial conditions.

RM requires a different treatment. The loading comprises a short impulse, which establishes finite $h_b$, $h_s$, $v_b$ and $v_s$ through rapid momentum deposition and early transition, followed by the zero-acceleration inertial regime of classical RM growth. All three buoyancy--drag models describe an established mixing layer and cannot reliably resolve the strong impulse and early transition. Integrating that stage from a small stationary perturbation would contaminate the hand-off state with out-of-scope error. Following Dimonte and Schneider \citep{dimonteschneider2000}, we therefore exclude the impulse from the model comparison and infer a common post-impulse state from the same experimental data.

On the falling flank of each impulse we define $t_*$ by $g(t_*)=0.5g_{\max}$, marked by the filled circles in figure~\ref{fig:4}(b). At this common relative threshold, most of the principal impulse has been delivered but the long tail has not fully decayed. This avoids selecting different dynamical stages merely because the pulses have different peaks and durations. Equation~\eqref{eq:4-1} gives $Z_*=Z(t_*)$, and $U_*=\int_0^{t_*}g(t)\,\mathrm{d}t$ gives the corresponding reference velocity. Local fits to the LEM width--displacement data near $Z_*$ yield $h_i^*$ and $(\mathrm{d}h_i/\mathrm{d}Z)_*$, from which $v_i^*=U_*(\mathrm{d}h_i/\mathrm{d}Z)_*$ for $i=b,s$. These four quantities define the post-impulse initial state.

From this state, all models use $\Delta Z=Z-Z_*$ and set $g_{\mathrm{model}}(t)=0$ for $t>t_*$, thereby solving only the inertial coasting regime. This computational definition does not imply that the experimental acceleration vanishes discontinuously at $t_*$; the measured pulse tail is simply excluded. The comparison thus isolates post-impulse predictive behaviour from uncertainty in the pulse tail and early transition. Table~\ref{tab:cases} lists the common initial states for the five RM cases.

\begingroup
\makeatletter\def\fps@table{h}\makeatother
\begin{table}
\centering
\small
\setlength{\tabcolsep}{4pt}
\caption{Common post-impulse initial states for the five RM cases.}
\label{tab:cases}
\begin{tabular}{rrrrrrr}
$R$ & $t_*$/ms & $Z_*$/cm & $h_b^*$/cm & $h_s^*$/cm & $v_b^*$/(cm\,s$^{-1}$) & $v_s^*$/(cm\,s$^{-1}$) \\
\midrule
1.57 & 15.862 & 9.0 & 0.24365 & 0.32133 & 49.008 & 46.825 \\
2.83 & 16.321 & 7.0 & 0.82739 & 0.68039 & 91.064 & 109.712 \\
12.1 & 12.988 & 4.0 & 0.41953 & 1.34412 & 87.911 & 305.386 \\
23.4 & 16.754 & 6.5 & 0.77667 & 1.47446 & 144.571 & 334.950 \\
49.1 & 15.998 & 6.0 & 0.79520 & 1.35464 & 295.030 & 388.091 \\
\end{tabular}
\end{table}
\endgroup

\subsection{Classical RT mixing: closure consistency and constant-acceleration evolution}\label{sec:4-2}

\subsubsection{RT growth coefficients and mixing-state branches}\label{sec:4-2-1}

\begin{figure}
\centering
\includegraphics[width=0.76\textwidth]{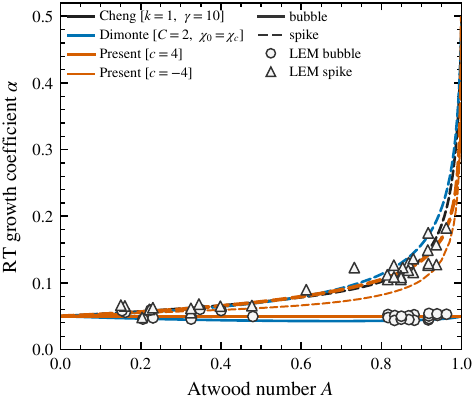}
\caption{Bubble and spike growth coefficients versus Atwood number in classical RT mixing. Curves show the present, Cheng and Dimonte models, and symbols denote LEM data; solid and dashed curves represent bubbles and spikes, respectively. The present model includes the principal LEM branch $c=4$ and a second mixing-state branch $c=-4$.}
\label{fig:3}
\end{figure}

Figure~\ref{fig:3} first tests the RT scaling closure. The bubble coefficient $\alpha_b=0.05$ is taken from LEM, so agreement on the bubble side confirms input consistency rather than constituting an independent prediction. The profile-shape parameter $c$ is selected from the LEM spike scaling $\alpha_s/\alpha_b\sim R^{0.33}$ across the measured density ratios. Applied to the profile family of \S~\ref{sec:3-2}, this relation selects $c=4$. It therefore identifies a mean-composition-profile state before the time histories are calculated; it is not fitted separately to those histories.

All three models recover $\alpha_s=\alpha_b$ as $A\to0$ and the free-fall spike limit $\alpha_s\to1/2$ as $A\to1$; their substantive differences therefore lie at intermediate density ratios. The $c=4$ branch captures the measured rise of the LEM spike coefficient with $A$, and the separation among the three closures is concentrated on the spike side in this range.

This difference clarifies the role of $c$. By controlling the interior profile shape, $c$ changes the bubble--spike width ratio through mass conservation. Over the LEM range, $c=4$ produces approximately $R^{0.33}$, whereas $c=-4$ gives the weaker $R^{0.22}$ branch. The branches share both endpoints but exhibit different asymmetry at intermediate density ratios. Mean-profile shape therefore affects not only the internal partition of the layer but also its macroscopic scaling, allowing the framework to represent mixing-state multiplicity at fixed density ratio.

\subsubsection{Mixing-width evolution under constant acceleration}\label{sec:4-2-2}

\begin{figure}
\centering
\includegraphics[width=\textwidth]{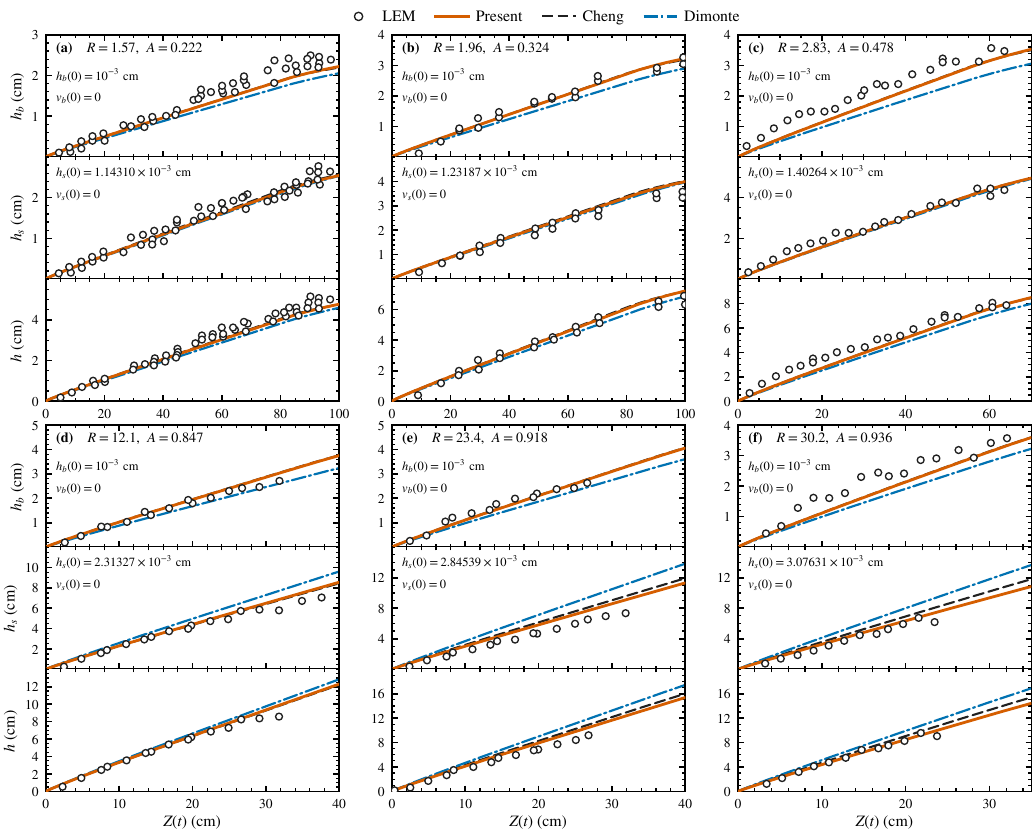}
\caption{Classical RT mixing widths versus reference displacement $Z(t)$ at six density ratios. From top to bottom, each group shows the bubble width $h_b$, spike width $h_s$ and total width $h$. Symbols denote LEM data; orange solid, black dashed and blue dash-dotted curves denote the present, Cheng and Dimonte models, respectively.}
\label{fig:5}
\end{figure}

Figure~\ref{fig:5} tests the resulting closure against quasi-constant-acceleration evolution at six density ratios. At low and moderate ratios ($R=1.57$, $1.96$ and $2.83$), all three models capture the nearly linear growth with $Z$. The present and Cheng results are close, and generally describe the bubble and total widths better than Dimonte. As $R$ increases, the bubble predictions remain similar while the models separate increasingly on the spike side.

The trend is clearest at high density ratio. At $R=12.1$, the present spike and total widths lie closer to the data than Dimonte's. At $R=23.4$ and $30.2$, Cheng and Dimonte substantially overpredict the late spike width, whereas the present closure suppresses this excess growth and correspondingly improves the total width. Some late overprediction remains.

Because $c$ controls both profile shape and width partition, case-specific adjustment could reduce these residuals. We deliberately retain one value for all cases. Figure~\ref{fig:5} therefore shows a cross-density-ratio prediction on a single profile branch, not a collection of posterior best fits. Without additional case-dependent freedom, explicit constraints on bubble--spike asymmetry and the high-density-ratio endpoint already improve the spike-dominated evolution substantially.

\subsection{Classical RM mixing: cross-problem prediction and post-impulse evolution}\label{sec:4-3}

\subsubsection{Cross-prediction of RM power-law exponents}\label{sec:4-3-1}

\begin{figure}
\centering
\includegraphics[width=0.76\textwidth]{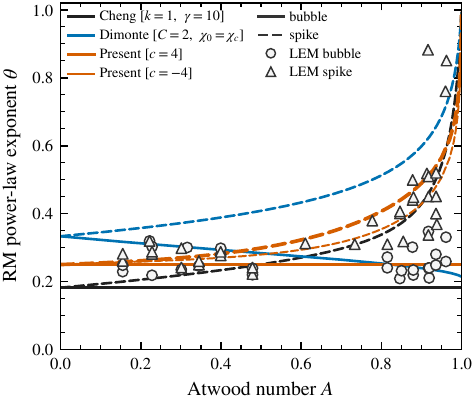}
\caption{Bubble and spike power-law exponents versus Atwood number in classical RM mixing. Curves and symbols are as in figure~\ref{fig:3}.}
\label{fig:6}
\end{figure}

Having tested the RT scaling closure, figure~\ref{fig:6} examines its extension to RM exponents. The bubble value $\theta_b=0.25$ is an LEM state input from \S~\ref{sec:3-2-5}; its agreement with experiment is therefore input consistency, not an independent prediction. In contrast, $c=4$ was selected entirely from RT spike scaling. RM spike data neither select the branch nor modify any coefficient. The predicted $\theta_s(A)$ is thus a direct cross-problem consequence of the RT scaling and mean-profile closure.

The predicted $\theta_s$ increases monotonically from the symmetric value $\theta_s=\theta_b$ at $A\to0$ to the ballistic limit $\theta_s\to1$ at $A\to1$. Between these endpoints, the curve captures the overall rise in the LEM spike exponent and passes through the main data cloud at moderate $A$. The scatter increases markedly at high $A$, but the prediction retains both the observed trend and the correct endpoint. This range therefore tests the global variation more meaningfully than pointwise ranking against individual measurements.

The models again separate mainly at moderate and high $A$. Cheng predicts both exponents too low, whereas Dimonte rises earlier and is generally high. The present model preserves the measured bubble baseline and gives a more balanced spike prediction. The $c=-4$ and $c=4$ branches share both endpoints but differ in the interior, mirroring the RT result in figure~\ref{fig:3}. A change of mean-profile state can therefore alter intermediate-density-ratio scaling in both RT and RM without changing either asymptotic limit.

These differences follow from the coefficient closures. Cheng infers bubble drag from the RT coefficient $\alpha_b$ and uses the same reduced relation to derive the RM exponent. With $\alpha_b=0.05$, it predicts $\theta_b\simeq0.182$, well below the LEM value $\simeq0.25$, and this low bubble baseline propagates to the spike. Dimonte assigns mean dynamical densities from a piecewise-linear density profile and uses one drag coefficient on both sides. With the fixed $C=2.0$, it predicts a bubble exponent above the LEM baseline and an earlier rise of the spike exponent, producing a high bias at moderate and large $A$.

The present model instead retains separate inertia, buoyancy and drag coefficients for each front. These are not six adjustable parameters; they provide the structural freedom required by distinct physical mechanisms. The inputs $\alpha_b$ and $\theta_b$ independently constrain the RT and RM bubble scalings, while similarity, mass conservation, the mean profile and endpoint asymptotics determine the remaining coefficients. The model can therefore preserve the measured RM bubble baseline and predict $\theta_s(A)$ without recalibration to any RM spike data.

\subsubsection{Mixing-width evolution after impulsive forcing}\label{sec:4-3-2}

\begin{figure}
\centering
\includegraphics[width=\textwidth]{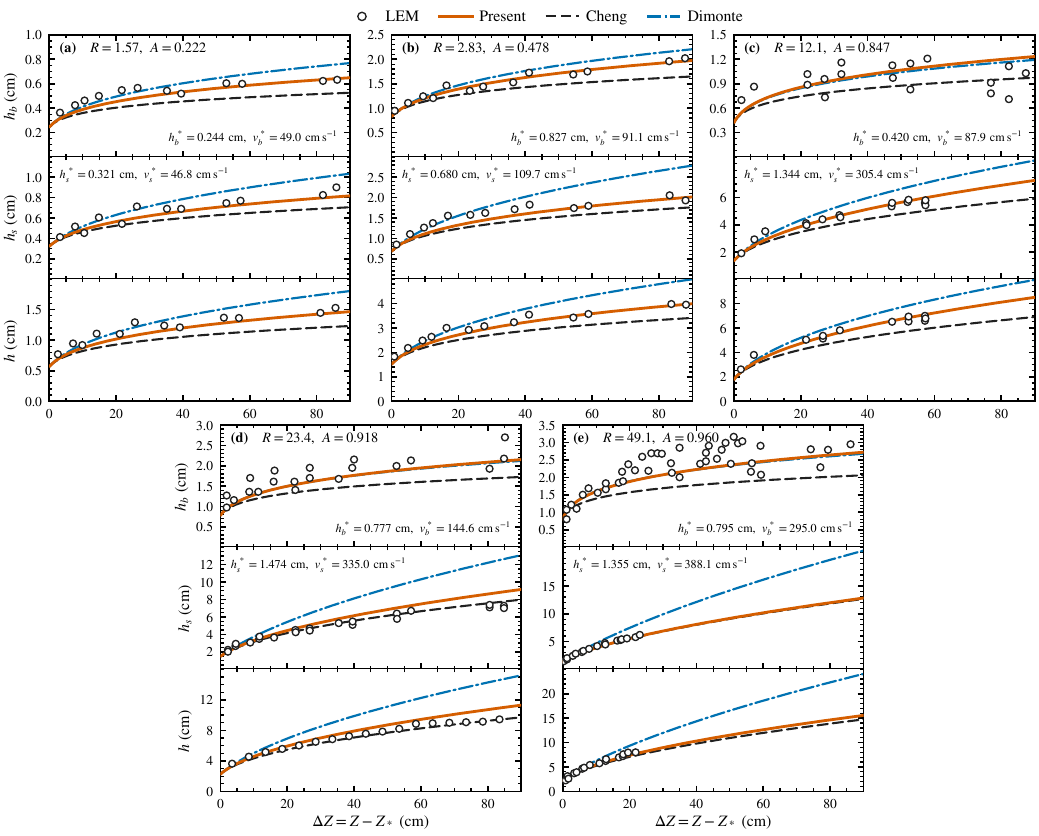}
\caption{Post-impulse RM widths versus $\Delta Z=Z-Z_*$ at five density ratios. From top to bottom, each group shows $h_b$, $h_s$ and $h$; symbols and curves are as in figure~\ref{fig:5}. The common initial widths and velocities are also reported.}
\label{fig:8}
\end{figure}

Figure~\ref{fig:8} next compares post-impulse width evolution at five density ratios. At $R=1.57$ and $2.83$, all models reproduce the basic increase of bubble, spike and total widths with $\Delta Z$, but differ in amplitude: Cheng is generally low, Dimonte high, and the present model gives a more balanced partition. At $R=12.1$, the present curves lie broadly between the benchmarks and follow all three measured widths well.

Because the three models use identical post-impulse widths and velocities, their subsequent separation directly reflects differences in drag closure and treatment of bubble--spike asymmetry. This separation is strongest at $R=23.4$ and $49.1$, again principally on the spike side. Relative to Dimonte, the present closure markedly suppresses excessive spike growth and maintains a coherent evolution of the two component widths and their sum. Cheng is close over some intervals but remains systematically low on the bubble side.

At $R=49.1$, spike measurements cover only a short $\Delta Z$ interval, so the curves beyond that range are model extrapolations. Taken together, figures~\ref{fig:6} and~\ref{fig:8} show that a mean-profile state selected by RT spike scaling captures both the global variation of the RM spike exponent and the principal post-impulse width evolution. Sharper discrimination at high $A$ will require measurements with broader coverage and less scatter.

\subsection{Transfer across variable-acceleration histories}\label{sec:4-4}

\subsubsection{Unsteady response on a fixed profile branch}\label{sec:4-4-1}

The preceding tests cover the two canonical limits of sustained buoyant forcing and post-impulse inertial evolution. In practice, acceleration often varies appreciably in both magnitude and direction. We therefore test whether the single closure fixed from the classical LEM cases transfers across distinct forcing histories.

\begin{figure}
\centering
\includegraphics[width=\textwidth]{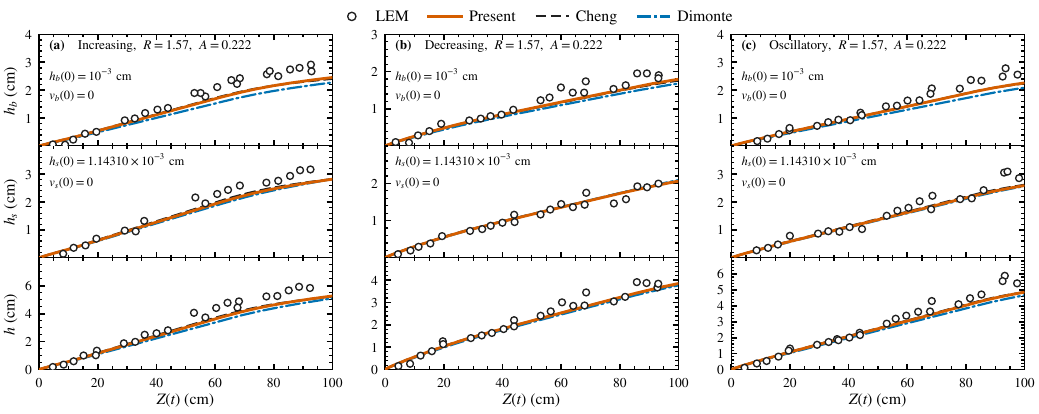}
\caption{Mixing-width evolution at $R=1.57$ under (a) increasing, (b) decreasing and (c) oscillatory acceleration. From top to bottom, each group shows $h_b$, $h_s$ and $h$; symbols and curves are as in figure~\ref{fig:5}.}
\label{fig:10}
\end{figure}

Figure~\ref{fig:10} compares increasing, decreasing and oscillatory accelerations at $R=1.57$. All models use the common initial conditions of \S~\ref{sec:4-1-3} and the complete measured history for each case. The present state parameters and dynamical coefficients remain fixed; only the external forcing changes.

All three models reproduce the continued growth under the changing histories, with the decreasing-acceleration case showing the closest agreement. The present and Cheng results are similar, while Dimonte generally gives slightly smaller bubble and total widths. Because $R=1.57$ is close to the symmetric endpoint, bubble--spike asymmetry is weak and the structural differences seen at moderate and high density ratios are not amplified. Nevertheless, the stable response of the present closure without retuning shows that it is not confined to quasi-constant acceleration.

The late experimental widths are slightly above the calculations in all three cases, most clearly for increasing and oscillatory acceleration. One possible source is the bubble coefficient $\alpha_b=0.05$, taken from classical LEM RT: $\alpha_b$ is state dependent, and a modest increase would raise both component widths. It is left unchanged here to test transfer of a single input set. A second possibility is genuine forcing-history dependence that cannot be represented by the instantaneous amplitude $g(t)$ alone. Although the complete acceleration history enters the equations, the inertia, buoyancy and drag coefficients depend only on density ratio and therefore assume quasi-instantaneous adjustment of the front structure to the external forcing. When the acceleration varies appreciably in time, the mixing-layer structure, entrainment and drag may respond over a finite adjustment time, causing the effective coefficients to depend on $\mathrm{d}g/\mathrm{d}t$ or, more generally, on the ratio of the loading time scale to the mixing-layer response time. Distinguishing variations in prescribed state inputs from genuinely unsteady memory effects, and quantifying their relative contributions to the present discrepancies, will require targeted experiments or high-fidelity simulations. Incorporating such finite-response and history effects into the coefficient closure is an important direction for future work.

\subsubsection{Aggregate quantitative errors across cases}\label{sec:4-4-2}

To complement the individual curves, table~\ref{tab:nrmse_summary} aggregates the normalised root-mean-square error (NRMSE) over the three classes of LEM time histories. For each class and width component, model curves are linearly interpolated to the experimental abscissae and all valid points are combined as
$\mathrm{NRMSE}=[\sum_j(h_{\mathrm{model},j}-h_{\mathrm{exp},j})^2/\sum_jh_{\mathrm{exp},j}^2]^{1/2}$
which normalises by the rms magnitude of the measured width while retaining all observations. The aggregate results confirm the trends from the individual figures. The present model gives the lowest total-width error for both constant-acceleration RT and post-impulse RM, principally because it constrains spike growth at moderate and high density ratio and thereby improves consistency among $h_b$, $h_s$ and $h$. The three models are closer for low-density-ratio variable-acceleration cases, where asymmetry is weak. The main gain is therefore the overall description of width partition and total growth across a broad density-ratio range.

{\color{black}
\begin{table}
\color{black}
\centering
\small
\caption{Aggregate NRMSE across three classes of LEM time histories. Boldface marks the smallest value for each case class and width component.}
\label{tab:nrmse_summary}
\begin{tabular}{llccc}
Case class & Model & $h_b$ & $h_s$ & $h$ \\
\midrule
Constant-acceleration RT & Present & 0.1423 & \textbf{0.1405} & \textbf{0.0977} \\
             & Cheng & \textbf{0.1399} & 0.1809 & 0.1125 \\
             & Dimonte & 0.2163 & 0.3177 & 0.1746 \\
\addlinespace
Post-impulse RM & Present & \textbf{0.1564} & 0.1191 & \textbf{0.0823} \\
          & Cheng & 0.2801 & \textbf{0.0917} & 0.0923 \\
          & Dimonte & 0.1590 & 0.4148 & 0.3048 \\
\addlinespace
Variable-acceleration RT & Present & \textbf{0.1629} & 0.1270 & 0.1316 \\
             & Cheng & 0.1632 & \textbf{0.1171} & \textbf{0.1251} \\
             & Dimonte & 0.2284 & 0.1387 & 0.1678 \\
\end{tabular}
\end{table}
}

\subsection{External assessment during the Nova deceleration stage}\label{sec:4-5}

\subsubsection{Experimental sequence and modelled stage}\label{sec:4-5-1}

The preceding tests all use the LEM facility and vary density ratio, forcing type and acceleration history within that experimental system. To assess whether the closure is facility dependent, we additionally consider the Nova laser experiments \citep{remington1995}. Their driving mechanism, materials, time scale and acceleration magnitude differ markedly from LEM, providing a more stringent cross-facility test.

After laser drive, the Nova targets pass through ablation, shock loading and, once the shock has traversed the mixing region, deceleration. Strong compressibility and shock effects place the first two stages outside the quasi-incompressible buoyancy--drag equations, so calculations begin at approximately $t_0=4\ \mathrm{ns}$ and cover only the subsequent deceleration. Let $g_{\mathrm{int}}(t)$ be the signed interface acceleration in the laboratory frame, positive in the original direction of motion. During deceleration $g_{\mathrm{int}}<0$, and we define the positive deceleration magnitude $g_d(t)\equiv-g_{\mathrm{int}}(t)>0$. The input $g_d(t)$ is digitised from the one-dimensional LASNEX interface history in figure 16 of Dimonte \citep{dimonte2000}; the total widths used for comparison are experimental Nova measurements \citep{remington1995,dimonte2000}. The measured width and growth rate at $t_0$ set the initial state. All dynamical coefficients remain those fixed from LEM, with no Nova-specific adjustment.

\subsubsection{Mixing-width prediction during deceleration}\label{sec:4-5-2}

\begin{figure}
\centering
\includegraphics[width=\textwidth]{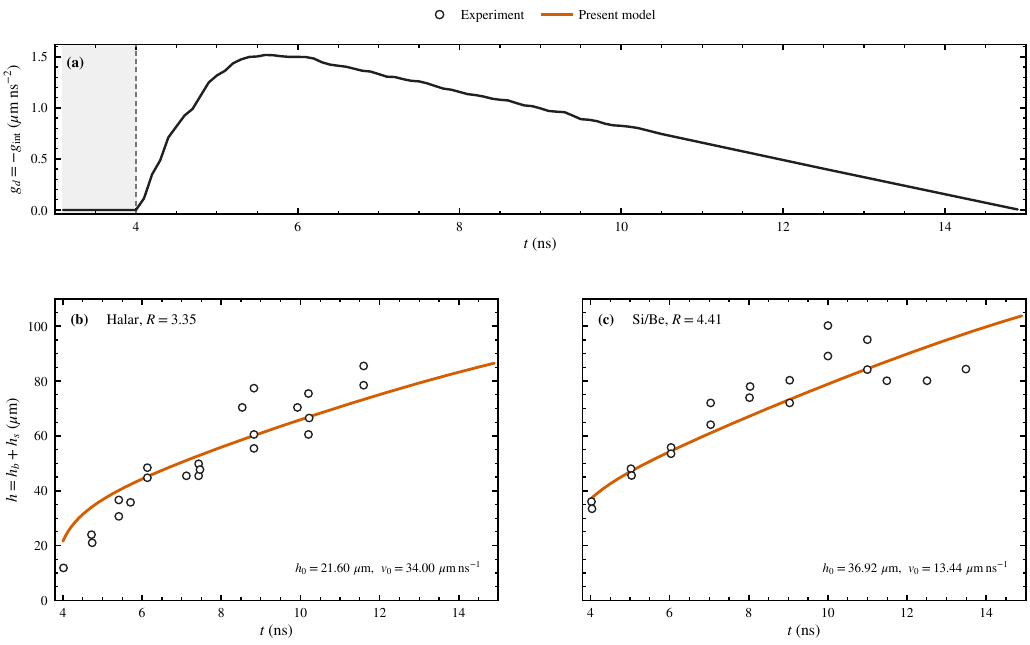}
\caption{Assessment during the Nova deceleration stage: (a) positive interface deceleration $g_d=-g_{\mathrm{int}}$, digitised from the LASNEX result in figure 16 of Dimonte \citep{dimonte2000}; the dashed line marks $t_0=4\ \mathrm{ns}$ and the shaded earlier interval is excluded from integration. Total widths are shown for (b) Halar and (c) SiBe targets. Symbols are experimental data, solid curves are the present model, and $h=h_b+h_s$.}
\label{fig:11}
\end{figure}

Figure~\ref{fig:11}(a) shows the positive interface deceleration $g_d$ used in the calculation; the shaded interval before $t_0$ is excluded. Panels (b,c) show the total width for Halar ($R\approx3.35$) and SiBe ($R\approx4.41$), testing evolution from the prescribed deceleration-stage initial state.

Within these boundaries, the model gives a consistent overall account of both targets. For Halar it slightly overpredicts the earliest point after initiation, passes through the main intermediate-time data cloud and falls below some high late measurements; the total-width RMSE is $7.30\ \mu\mathrm{m}$ and the relative RMSE is $0.147$. For SiBe it follows the early growth, lies below several measurements over $7$--$10\ \mathrm{ns}$ and exceeds the final two relatively low points; the corresponding values are $8.47\ \mu\mathrm{m}$ and $0.119$. Despite appreciable scatter, the curves traverse the principal data range and reproduce the growth trend and overall magnitude without an additional decompression correction.

Dimonte also applied the spanwise-homogeneous model to Nova \citep{dimonte2000}. That calculation began earlier after shock passage, partitioned the first measured total width into bubble and spike components, estimated their velocities from RM growth, nonlinear saturation and transmitted-shock speed, and then applied both the LASNEX deceleration and a foam-decompression correction. The present calculation instead begins from the measured state at $t_0=4\ \mathrm{ns}$, retains the LEM closure and applies only the subsequent deceleration, without a multiplicative decompression factor. Because the initial stages, state construction and physical corrections differ, the published Dimonte result is a methodological and trend reference rather than a like-for-like accuracy benchmark for figure~\ref{fig:11}.

The facility, materials, time scale and acceleration history all change between LEM and Nova, yet the dynamical closure is unchanged. Its ability to reproduce the principal deceleration-stage evolution of both targets demonstrates cross-facility transfer. Because the state at $t_0$ is taken from Nova data and the ablation, shock-loading and early-transition stages are excluded, figure~\ref{fig:11} tests evolution after a prescribed initial state, not an end-to-end prediction from laser turn-on. It is a stage-specific external assessment that also delineates the strongly compressible and shock-driven regimes beyond the present model.

\section{Discussion}
\label{sec:5}

\subsection{Connecting local front dynamics to global mass conservation}
\label{sec:5-1}

At the heart of the present modelling advance is the integration of local front dynamics and global mass conservation within a single low-order framework. Although the interior is a statistically averaged turbulent region, its bubble and spike widths are set directly by the motion of the two fronts. Separate equations for representative front structures therefore give inertia, buoyancy and drag explicit local mechanical carriers. The fronts cannot, however, evolve independently of the layer they bound: their width ratio must satisfy global mass conservation under the mean-composition profile. Local equations establish the dynamical structure, while global conservation couples the two sides.

In the local construction, the bubble net inertia is the sum of added-mass and entrainment contributions, whereas on the spike side entrainment subtracts from added mass. This separates the non-mixing inertial coupling represented by added mass from the mixed-mass contribution represented by entrainment, and leads to a volume-ratio interpretation of buoyancy through normalisation by the coherent and effective inertial volumes. The low-order equations ultimately close the net combinations rather than separately measuring added-mass and entrainment coefficients. The decomposition retains mechanistic traceability without introducing independent quantities that current width data cannot identify.

\subsection{Coefficient closure, density-ratio dependence and asymptotic consistency}
\label{sec:5-2}

Unlike models that impose a common drag coefficient or characteristic scale, the present formulation retains separate inertia, buoyancy and drag coefficients with distinct density-ratio dependence. Retaining six coefficients does not mean prescribing six arbitrary functions. The bubble inertia coefficient is fixed from the idealised geometric input $r(0)$ at $A=0$ and held constant; the spike inertia is constrained by the symmetric and vacuum endpoints; and buoyancy and drag are inverted from RT/RM similarity. The architecture preserves intrinsic bubble--spike asymmetry while placing every effective coefficient under an explicit physical constraint.

Endpoint limits are built directly into the coefficient closure. The two sides become symmetric at $A=0$. As $A\to1$, the density ahead of the spike vanishes, taking both density-weighted added inertia and drag to zero and recovering a free-falling RT spike and a ballistic RM spike. Equation~\eqref{eq:3-24}, which connects the RT and RM spike scalings, is an additional cross-problem hypothesis rather than a unique consequence of the endpoints; asymptotic matching fixes only its vacuum-limit exponent $p_1$. \redtext{The same endpoint also clarifies why the large isolated value of the spike drag coefficient is not pathological: although $C_D^s$ grows to approximately $0.9\lambda+0.5$, the density-weighted drag $C_D^s\rho_1$ entering the dynamics vanishes and preserves the correct physical limit.}

The quadratic form~\eqref{eq:3-25} is the lowest-order even function satisfying symmetry, endpoint values and monotonicity, but it is not the only admissible interpolation. Appendix~\ref{app:mapping-sensitivity} tests an alternative quartic form subject to the same endpoint constraints. The two mappings produce $\theta_s$ curves that differ by at most about $0.03$ while preserving the same monotonic trend and endpoint limits. Within this physically constrained family of low-order functions, the RM cross-prediction exhibits low sensitivity to the specific interpolation order, and both its overall trend and principal physical conclusions remain robust.

\subsection{Cross-RT/RM prediction, state multiplicity and relation to earlier models}
\label{sec:5-3}

The formulation also connects two earlier modelling routes. \redtext{The nonlinear-profile model of Zhang et al.\ \citep{zhang2016} centres on a statistically averaged layer and a quasi-momentum conservation constraint; Li and Zhang \citep{li2025} subsequently related bubble and spike widths using a mean-composition profile and exact mass conservation.} The present work returns to representative front structures, writes a dynamical equation for each side, and uses that mean profile and conservation relation to constrain the local coefficients. Conservation acts on the integral relation between penetration widths and composition, rather than assuming equal penetration volumes of light and heavy fluid.

The parameter $c$ is a state coordinate of the mean-profile family that represents the multiplicity arising from sensitivity to initial perturbations, miscibility and other state variables. Quadratic coefficients and power-law exponents are not unique at fixed density ratio, as shown by LEM, earlier experiments and simulations. Different values of $c$ select different profiles, which mass conservation maps into different bubble--spike width ratios. Thus $c$ is not a pointwise correction to a time history but a coordinate selecting a family of mixing states. The framework propagates that multiplicity once $c$ is given, although it does not yet predict $c$ from the initial perturbation.

On the LEM branch, the RT spike coefficient selects the mean profile and the RM spike exponent follows from the cross-problem relation. This provides a stricter joint test of bubble and spike behaviour than matching total width alone. Once the branch is fixed, the remaining coefficients describe constant, impulsive and variable acceleration without retuning. Under the benchmark settings used here, the most pronounced improvement occurs for high-density-ratio spikes.

\subsection{Scope and future directions}
\label{sec:5-4}
The model closes effective dynamical quantities identifiable from mixing-width data. Entrainment and added mass enter through a net inertia coefficient; richer measurements of local velocity, composition flux or front geometry could separate their individual contributions. The value $r(0)=3$ provides a definite symmetric-endpoint reference from an idealised hemispherical-cap/cylindrical geometry. More detailed front measurements could update $r(0)$ and its associated coefficients using the same volume definitions.

Further development concerns both state inputs and the RT--RM mapping. The parameters $\alpha_b$, $\theta_b$ and $c$ describe bubble growth and mean-profile state and are fixed globally from LEM. A quantitative link to initial-perturbation spectra, miscibility and loading history would permit direct prediction of the mixing state from initial conditions. Equation~\eqref{eq:3-24} is already constrained by the symmetric and vacuum endpoints and by the mapping-sensitivity test; additional independent RM data could refine its finite-$A$ form.

The present front equations address an established mixing layer under quasi-incompressible conditions. Validation therefore covers constant- and variable-acceleration RT, the post-impulse inertial stage of RM, and Nova deceleration after a prescribed initial state. Extending the model to linear instability, momentum deposition during a strong impulse, ablation, shock propagation and strongly compressible stages will require coupling the current front dynamics to models of early transition, shock dynamics or compressible flow \citep{cao2025double}.
\section{Conclusions}
\label{sec:6}

We have developed a physically constrained buoyancy--drag model for macroscopic width prediction in wide-density-ratio RT and post-impulse RM mixing. The model connects local front dynamics to global mass conservation: separate bubble and spike equations distinguish coherent mass, added mass and entrainment within a Newtonian framework, while a mean-composition profile and mass conservation establish the quantitative relation between the two penetration widths. Bubble--spike asymmetry, density-ratio dependence and distinct forcing mechanisms are thereby brought into a single low-order framework.

Rather than imposing shared or fixed empirical coefficients, the formulation retains separate inertia, buoyancy and drag coefficients on the two sides and permits distinct density-ratio dependence. Given bubble-side scalings, a mean-profile state and the symmetric-endpoint geometry, all six effective coefficients are jointly determined by RT/RM similarity, global mass conservation and endpoint asymptotics; they are not independent case-by-case fitting parameters. By construction, the closure recovers bubble--spike symmetry at low Atwood number, free-fall RT spike scaling and ballistic RM spike scaling at the high-density-ratio limit. The mean-profile parameter $c$ further links internal mixing state to the macroscopic width partition, enabling the same framework to represent distinct branches of asymmetric growth.

Once the profile state is selected from the LEM RT spike scaling, the model cross-predicts the rise of the RM spike exponent with Atwood number without recalibration to RM spike data. Tests across density ratio for constant-acceleration RT and post-impulse RM show that the closure suppresses excessive high-density-ratio spike growth and improves consistency among bubble, spike and total widths. The same state inputs and coefficients respond stably to increasing, decreasing and oscillatory accelerations and transfer to the deceleration stage of the Nova laser experiments. Together, these results form a validation chain from RT scaling closure and RM cross-prediction to unsteady forcing and cross-facility assessment.

The principal contribution of this work lies not only in improving width predictions for specific conditions but, more fundamentally, in establishing a complete closure route that begins with local dynamics, couples the fronts through global conservation and constrains the model coefficients by asymptotic limits. The framework reveals the intrinsic connection between mean mixing state, bubble--spike asymmetry and macroscopic growth scaling. It thereby provides a unified low-order modelling methodology that is physically interpretable and asymptotically consistent, accommodates multiple mixing states, and supports cross-problem prediction of wide-density-ratio RT and post-impulse RM mixing.

\begin{acknowledgments}
\textbf{Declaration of Interests.} The authors report no conflict of interest.
\end{acknowledgments}

\begin{acknowledgments}
The authors acknowledge financial support from the National Natural Science Foundation of China (grant nos. 12588301, 12532013 and 12672283) and the National Key Laboratory of Computational Physics (grant no. 6142A05240201).
\end{acknowledgments}

\appendix

\section{First-order asymptotic expansion at the vacuum limit}\label{app:asymptotic}

{\color{black}
Let $\epsilon=1-A\to0^+$, so that $R=(1+A)/(1-A)=2/\epsilon-1$. Equation~\eqref{eq:3-11} gives $C_I^s=2C_I^b(0)\epsilon+O(\epsilon^2)$. Using $\Theta_\alpha(1-\epsilon)=1-\lambda\epsilon+O(\epsilon^2)$, equation~\eqref{eq:3-13} becomes
\begin{equation}
\alpha_s=\frac12-\left(\frac12-\alpha_b\right)\lambda\epsilon+O(\epsilon^2).
\label{eq:A1}
\end{equation}
Define $q=(1-\theta_b)p_1\lambda$. Since $p(1-\epsilon)=p_1+O(\epsilon)$ and $\ln\Theta_\alpha=-\lambda\epsilon+O(\epsilon^2)$, equations~\eqref{eq:3-23}--\eqref{eq:3-24} give
\begin{equation}
\theta_s=1-q\epsilon+O(\epsilon^2).
\label{eq:A2}
\end{equation}

Substituting equations~\eqref{eq:A1}--\eqref{eq:A2} into~\eqref{eq:3-8}, and noting that
\begin{equation}
\frac{C_I^s+R}{R+1}=1-\frac{\epsilon}{2}+O(\epsilon^2),
\end{equation}
gives
\begin{equation}
C_B^s=1+\left[-\frac12-2\left(\frac12-\alpha_b\right)\lambda+2q\right]\epsilon+O(\epsilon^2).
\label{eq:A4}
\end{equation}
Requiring $C_B^s=1+O(\epsilon^2)$ gives $q=(1/2-\alpha_b)\lambda+1/4$ and hence equation~\eqref{eq:3-26}. Finally, substitution into~\eqref{eq:3-9} yields
\begin{equation}
\lim_{A\to1}C_D^s=2q=(1-2\alpha_b)\lambda+\frac12,
\end{equation}
which is equation~\eqref{eq:3-27}. This first-order expansion is used only for vacuum-limit matching; all finite-$A$ curves in the main text use numerical integration of the complete mean profile rather than this local expansion.
}

\section{\redtext{Sensitivity test of the RT--RM mapping form}}\label{app:mapping-sensitivity}

{\color{black}
To test whether the lowest-order quadratic choice controls the RM spike result, we substitute $p_2(A)=1-(1-p_1)A^2$ and $p_4(A)=1-(1-p_1)A^4$ separately into equation~\eqref{eq:3-24}. Both are even, satisfy $p(0)=1$ and $p(1)=p_1$, and remain monotone. Figure~\ref{fig:p-sensitivity} shows that the quartic form gives a slightly smaller $\theta_s$ at moderate and high Atwood number. The maximum difference is about $0.03$, while both retain the symmetric endpoint at $A=0$, the ballistic endpoint at $A=1$ and monotonic growth between them. The interpolation introduces a limited quantitative uncertainty but does not alter the principal trend or endpoint conclusions.

\begin{center}
\begin{minipage}{0.86\textwidth}
\centering
\includegraphics[width=\textwidth]{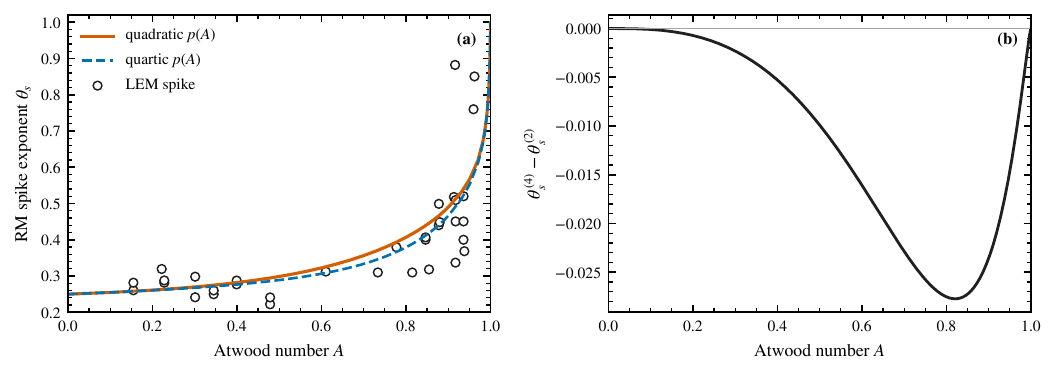}
\captionof{figure}{Sensitivity to the RT--RM mapping form: (a) RM spike exponents from quadratic and quartic $p(A)$; (b) difference between the predictions. Symbols denote LEM spike data.}
\label{fig:p-sensitivity}
\end{minipage}
\end{center}
}

\bibliographystyle{jfm}
\bibliography{references}
\end{document}